\documentclass[sigconf]{acmart}
\AtBeginDocument{%
  }

\setcopyright{acmlicensed}
\copyrightyear{2018}
\acmYear{2018}
\acmDOI{XXXXXXX.XXXXXXX}
\acmConference[Conference acronym 'XX]{Make sure to enter the correct
  conference title from your rights confirmation email}{June 03--05,
  2018}{Woodstock, NY}
\acmISBN{978-1-4503-XXXX-X/2018/06}

\usepackage[utf8]{inputenc} 
\usepackage[T1]{fontenc}    
\usepackage{pifont}

\usepackage{graphicx}
\usepackage{xcolor}
\usepackage{todonotes}
\usepackage[table,xcdraw]{xcolor}

\usepackage{amssymb}
\usepackage{threeparttable}
\usepackage{multirow}
\usepackage{amsmath}
\usepackage{bm}
\usepackage{float}
\usepackage{amsthm}
\usepackage{soul}
\usepackage[noend]{algpseudocode}
\usepackage{enumitem} 
\usepackage[subrefformat=parens,labelformat=parens]{subfig}

\usepackage{xspace}

\usepackage{multirow}

\usepackage[normalem]{ulem}
\usepackage{soul}
\usepackage{booktabs}

\definecolor{citecolor}{RGB}{34,139,34}
\definecolor{mydarkblue}{rgb}{0,0.08,1}
\definecolor{mydarkgreen}{rgb}{0.02,0.6,0.02}
\definecolor{mydarkred}{rgb}{0.8,0.02,0.02}
\definecolor{mydarkorange}{rgb}{0.40,0.2,0.02}
\definecolor{mypurple}{RGB}{111,0,255}
\definecolor{myred}{rgb}{1.0,0.0,0.0}
\definecolor{mygold}{rgb}{0.75,0.6,0.12}
\definecolor{myblue}{rgb}{0,0.2,0.8}
\definecolor{mydarkgray}{rgb}{0.,0.2,0.2}

\definecolor{lightred}{RGB}{255,235,235}
\definecolor{lightgreen}{RGB}{235,255,235}
\definecolor{lightblue}{RGB}{235,235,255}
\definecolor{lightcyan}{RGB}{235,255,255}
\definecolor{lightmagenta}{RGB}{255,235,255}
\definecolor{lightyellow}{RGB}{255,255,235}

\definecolor{qxkcolor}{RGB}{215,235,255}
\definecolor{softmaxcolor}{RGB}{230,235,255}
\definecolor{probxvcolor}{RGB}{255,255,235}

\definecolor{topkcolor}{RGB}{255,235,235}
\definecolor{zecolor}{RGB}{255,255,235}
\definecolor{dynacolor}{RGB}{235,255,255}

\definecolor{reviewcolor}{RGB}{0,0,200}

\theoremstyle{plain}

\theoremstyle{definition}

\algdef{SE}[DOWHILE]{Do}{doWhile}{\algorithmicdo}[1]{\algorithmicwhile\ #1}%

\newcommand{\squishlist}{
 \begin{list}{$\bullet$}
  { \setlength{\itemsep}{0pt}
     \setlength{\parsep}{3pt}
     \setlength{\topsep}{3pt}
     \setlength{\partopsep}{0pt}
     \setlength{\leftmargin}{1.5em}
     \setlength{\labelwidth}{1em}
     \setlength{\labelsep}{0.5em} } }
     
\newcommand{\squishend}{
  \end{list}  }     

\newcommand{\name}{\texttt{DPE-Rec}\xspace}

\begin{document}


\title{\name: \underline{D}iversity-\underline{P}reserving and \underline{E}fficient Diffusion Language Model \underline{Rec}ommendation for Bundle Construction}



\author{Pingchuan Ma}
\correspondingauthor
\affiliation{%
  \institution{Arizona State University}
  \city{Tempe}
  \state{Arizona}
  \country{USA}}
\email{pingchua@asu.edu}

\author{Bohan Jiang}
\affiliation{%
  \institution{Arizona State University}
  \city{Tempe}
  \state{Arizona}
  \country{USA}}
\email{bjiang14@asu.edu}

\author{Chengshuai Zhao}
\affiliation{%
  \institution{Arizona State University}
  \city{Tempe}
  \state{Arizona}
  \country{USA}}
\email{czhao93@asu.edu}

\author{Dawei Li}
\affiliation{%
  \institution{Arizona State University}
  \city{Tempe}
  \state{Arizona}
  \country{USA}}
\email{daweili5@asu.edu}

\author{Zhen Tan}
\affiliation{%
  \institution{Stevens Institute of Technology}
  \city{Hoboken}
  \state{New Jersey}
  \country{USA}}
\email{ztan12@stevens.edu}

\author{Huan Liu}
\affiliation{%
  \institution{Arizona State University}
  \city{Tempe}
  \state{Arizona}
  \country{USA}}
\email{huanliu@asu.edu}







\renewcommand{\shortauthors}{Anonymous et al.}

\begin{abstract}
Generative bundle construction completes a partially observed bundle by generating its missing items from a large product catalog.
Compared to autoregressive LLMs, diffusion LLM emerges as a natural fit for this unordered generation task.
Their bidirectional attention provides an inductive bias better aligned with unordered bundle construction.
Iterative denoising can reveal multiple tokens per step, potentially reducing inference latency.
However, directly applying DLMs to this task presents two key challenges:
\ding{202}~Despite its flexible reveal order, confidence-guided decoding tends to postpone semantic-fork decisions until refinement tokens constrain their choices, reducing bundle diversity.
\ding{203}~Revealing more tokens per step can compromise recommendation quality, stifling the potential for faster inference.
We introduce \name, a diffusion framework for diversity preservation and efficient recommendation.
\name preserves diversity by introducing semantic-fork-aware GRPO, which prioritizes semantic-fork decisions via autoregressive rollouts. 
It improves efficiency through pseudo-trajectory distillation, which concentrates learning from the vast random corruption space onto the compact trajectory-induced subspace visited by high-quality decoding, unlocking reliable multi-token decoding.
Together, these components enable diverse and accurate bundle construction with substantially fewer denoise steps. 
Experiments on multiple bundle construction benchmarks show that \name improves recommendation performance by 63\% on average, and its fast variant achieves 25.6$\times$ faster inference on average than existing generative recommendation arts, with minimal loss in recommendation quality.
\end{abstract}



\received{20 February 2007}
\received[revised]{12 March 2009}
\received[accepted]{5 June 2009}

\maketitle

\section{Introduction}

Bundle construction completes a partially observed bundle by generating a coherent set of missing items from a large product catalog \cite{chang2021bundle,jeon2023accurate,ma2024leveraging,liu2025fine,tu2026discrete}. Widely adopted in e-commerce \cite{zhu2014bundle} and online services such as music and video streaming \cite{vall2017music,ferraro2018automatic}, it has become a foundational recommendation task whose unordered output must capture bundle intent, cross-item compatibility, and set diversity while supporting efficient generation at catalog scale.

Existing approaches broadly follow discriminative and generative paradigms. ID based sequential models\cite{hidasi2015session,tang2018personalized,yuan2019simple,kang2018self,sun2019bert4rec,li2020time,zhou2020s3,zhai2024actions} capture user preferences from interaction sequences through recurrent, convolutional, or attention-based architectures. However, their reliance on sparse item identities limits semantic sharing across related items. Representation enhanced discriminative models\cite{zhang2019feature,petrov2024recjpq,ma2022crosscbr,ma2024multicbr,ma2024leveraging} alleviate this limitation by incorporating item features, multiple interaction views, or compositional representations. Nevertheless, they still retrieve or rank catalog candidates, implicitly modeling dependencies among jointly selected items only. Semantic ID (SID) based generative models\cite{rajputrecommender,si2024generative,jin2023language,zheng2024adapting,wang2024eager,li2026unigrec,liu2026diffgrm,tu2026discrete} instead represent each item with a sequence of discrete tokens and directly generate target identifiers. Autoregressive generation further captures dependencies among generated items through sequential conditioning. Yet its fixed left-to-right generation order imposes an artificial ordering on an inherently unordered set, resulting in an inductive bias poorly aligned with bundle construction.

\begin{figure}
    \centering
    \subfloat[]{\includegraphics[width=0.48\columnwidth]{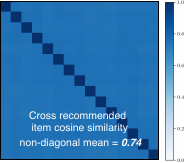}
    \label{fig:motivation_high_sim}
    }
    \hspace{0pt}
    \subfloat[]{\includegraphics[width=0.42\columnwidth]{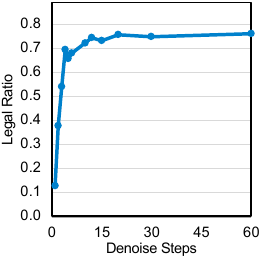}
    \label{fig:motivation_tradeoff}
    }
    \vspace{-12pt}
    \caption{
    (a) High similarity among items generated by the DLM bundle recommender;
    (b) Conservative decoding with more denoising steps produces legal SIDs, whereas aggressive decoding with fewer steps produces illegal SIDs.
    }
\label{fig:motivation}
    \vspace{-25pt}
\end{figure}

Diffusion language models (DLMs) offer a promising alternative by constructing bundle SIDs through iterative masked denoising \cite{tu2026discrete}. Their bidirectional attention provides an inductive bias better aligned with unordered bundle construction. Iterative denoising can reveal multiple tokens per step, potentially reducing inference latency. In confidence-guided decoding, the model ranks remaining masked positions by confidence and reveals one or more of the highest-confidence tokens per round. The ranking determines the reveal order, while the number of revealed tokens controls parallelism. Yet realizing these advantages presents two critical challenges.
\ding{202}~Despite its flexible reveal order, confidence-guided decoding tends to postpone semantic-fork decisions. The first SID token selects a coarse semantic region of the catalog and typically has higher uncertainty than later refinement tokens. Confidence ranking therefore prioritizes high-confidence refinement tokens, forcing the semantic fork to accommodate these earlier decisions rather than explore alternative semantic branches. Figure~\ref{fig:motivation_high_sim} shows high pairwise cosine similarity among fifteen generated items, illustrating the resulting loss of bundle diversity.
\ding{203}~Revealing more tokens per step can compromise recommendation quality, stifling the potential for faster inference. Because simultaneously revealing dependent SID tokens relies on a bold conditional independence approximation, individually plausible predictions can combine into illegal SIDs. Figure~\ref{fig:motivation_tradeoff} shows that aggressive parallel decoding reduces latency but sacrifices SID validity, whereas conservative decoding requires many more denoising rounds.
To overcome these limitations, we introduce \name, a diffusion framework for diversity preservation and efficient inference. Semantic-fork-aware Group Relative Policy Optimization (GRPO) \cite{shao2024deepseekmath} preserves diversity through fork-first autoregressive rollouts and higher sampling temperatures at semantic forks. Pseudo-trajectory distillation improves efficiency by concentrating learning from the vast random corruption space onto the compact trajectory-induced subspace visited by high-quality decoding, unlocking reliable multi-token decoding. Together, these designs enable diverse and accurate bundle construction with substantially fewer denoising rounds.

Our main contributions are summarized as follows:

\begin{itemize}
    \item We identify two fundamental limitations of diffusion bundle recommenders. Confidence-guided arbitrary-order decoding prioritizes refinement tokens over semantic forks, restricting bundle diversity, while aggressive parallel decoding relies on a bold conditional independence approximation that compromises recommendation quality.
    
    \item We propose semantic-fork-aware GRPO to preserve bundle diversity by confronting semantic forks while their uncertainty remains intact. It combines fork-first autoregressive rollouts with level-specific sampling temperatures, without altering the bidirectional denoiser or its parallel decoding capability.

    \item We introduce pseudo-trajectory distillation to concentrate denoising learning on the compact subspace induced by high-quality, sequential, confidence-guided trajectories. This enables the model to preserve dependencies among SID tokens while resolving multiple tokens per denoising round, substantially improving inference efficiency without sacrificing recommendation quality.

    \item Experiments on multiple bundle construction benchmarks show that \name improves recommendation performance by 63\% on average, while its fast variant achieves 25.6 times faster inference on average than existing generative recommenders, with minimal loss in recommendation quality.
\end{itemize}
\section{Preliminary}
\subsection{Diffusion Language Model}
\begin{figure*}[t]
    \centering
    \includegraphics[width=1\textwidth]{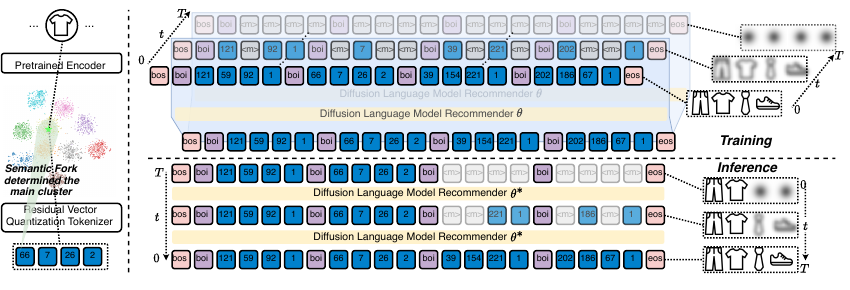}
    \caption{Overview of diffusion language model-based bundle recommendation. Item representations from a pretrained encoder are discretized into semantic identifiers through residual vector quantization, where successive SID tokens encode progressively finer semantics, and the first token primarily determines a coarse item cluster. During training, each token is independently replaced with $<\mathrm{m}>$ with probability $t$, and the denoiser reconstructs the clean sequence using token-wise cross entropy over masked positions. During inference, the SIDs of observed items remain visible while the missing item positions are initialized with $<\mathrm{m}>$. Iterative denoising then reveals one or more tokens at each round until the bundle is completed.}
    \label{fig:dlm_recommender}
\end{figure*}
Diffusion language models generate discrete sequences through iterative masked denoising. Given a clean sequence $\mathbf{x}_{0} = [x_{0,1}, \ldots, x_{0,L}]$, the forward process independently replaces each token with mask token $<\mathrm{m}>$ according to a masking ratio $t \in [0,1]$:

\begin{equation}
q(x_{t,k} \mid x_{0,k}) =
\begin{cases}
<\mathrm{m}>, & \text{with probability } t, \\
x_{0,k}, & \text{with probability } 1-t,
\end{cases}
\end{equation}

where $k$ denotes the token position. A neural network $p_{\theta}(\mathbf{x}_{0} \mid \mathbf{x}_{t})$ predicts the original tokens at masked positions.

During inference, generation starts from a fully masked or partially observed sequence. At each round, the model predicts all masked positions and reveals a subset of tokens according to a decoding policy, commonly based on prediction confidence. This process updates $\mathbf{x}_{t}$ to $\mathbf{x}_{t-\Delta t}$ until all tokens are revealed. Autoregressive generation is a special case that always reveals the leftmost masked token.

The model is trained by minimizing the negative evidence lower bound, which reduces to a weighted cross-entropy objective:

\begin{equation}
\mathcal{L}_{\mathrm{DLM}}(\theta)
=
-\mathbb{E}_{t \sim \mathcal{U}[0,1],\, \mathbf{x}_{t} \sim q(\mathbf{x}_{t} \mid \mathbf{x}_{0})}
\left[
\frac{1}{t}
\sum_{k=1}^{L}
\mathbb{I}[x_{t,k} = <\mathrm{m}>]
\log p_{\theta}(x_{0,k} \mid \mathbf{x}_{t})
\right].
\end{equation}

\subsection{Group Relative Policy Optimization}
\label{sec:prelim_grpo}
GRPO is a reinforcement learning algorithm for autoregressive policies, whose explicit token factorization enables tractable likelihood estimation and direct credit assignment without a learned value function. Given a policy $\pi_{\theta}(o_k \mid o_{<k}, \mathbf{c})$, GRPO samples a group of $G$ outputs $\{o_i\}_{i=1}^{G}$ from the old policy $\pi_{\theta_{\mathrm{old}}}$ and normalizes their rewards within the group:

\begin{equation}
\hat{A}_{i}
=
\frac{r(o_i)-\mu_G}{\sigma_G},
\end{equation}

where $\mu_G$ and $\sigma_G$ are the mean and standard deviation of the group rewards. GRPO then optimizes a clipped surrogate objective with KL regularization:

\begin{equation}
\mathcal{J}(\theta)
=
\mathbb{E}
\left[
\frac{1}{G}
\sum_{i=1}^{G}
\frac{1}{|o_i|}
\sum_{k=1}^{|o_i|}
\left[
\min
\left(
\rho_{i,k}\hat{A}_{i},
\operatorname{clip}
\left(
\rho_{i,k},
1-\epsilon,
1+\epsilon
\right)
\hat{A}_{i}
\right)
-
\beta D_{\mathrm{KL}}
\right]
\right],
\end{equation}

where the token-level importance ratio is

\begin{equation}
\rho_{i,k}
=
\frac{
\pi_{\theta}
\left(
o_{i,k}
\mid
o_{i,<k},\mathbf{c}
\right)
}{
\pi_{\theta_{\mathrm{old}}}
\left(
o_{i,k}
\mid
o_{i,<k},\mathbf{c}
\right)
}.
\end{equation}
\section{Understanding the Quality and Efficiency Limitations}

Before introducing \name, we examine two limitations of diffusion-based bundle construction. \ding {202} ~ confidence-guided arbitrary-order decoding postpones uncertain early SID decisions, reducing bundle diversity and producing highly similar recommendations. \ding{203}~parallel decoding creates an inherent quality and efficiency tradeoff, forcing a choice between unreliable SID generation and excessive inference latency.

\subsection{The Flexibility Trap in Diffusion Recommenders}
\label{sec:flexibility_trap}
\paragraph{Arbitrary Order Decoding Restricts Recommendation Exploration.}
\begin{figure}
    \centering
    \subfloat[]{\includegraphics[width=0.48\columnwidth]{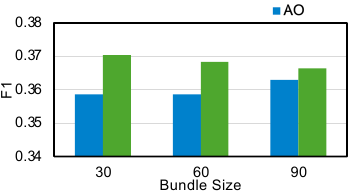}
    \label{fig:ar_over_ao}
    }
    \hspace{0pt}
    \subfloat[]{\includegraphics[width=0.48\columnwidth]{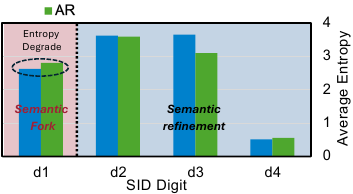}
    \label{fig:sfork_ereduce}
    }
    \caption{
    (a) For the same model, autoregressive decoding consistently outperforms the arbitrary order decoding;
    (b) Mean entropy of each SID position at the step it is decoded under AR and AO. AR preserves semantic fork entropy, whereas AO causes entropy degradation by resolving refinement tokens first.
    }
    \label{fig:flex_trap_explain}
\end{figure}

The flexibility of arbitrary order decoding does not necessarily translate into broader recommendation exploration. To examine this effect, we evaluate the same diffusion recommender checkpoint under two inference orders while keeping the remaining decoding and evaluation settings unchanged. The conventional policy reveals tokens according to prediction confidence and therefore allows SIDs to be generated in an arbitrary order, which we denote as AO. In contrast, autoregressive decoding, denoted as AR, does not introduce a separate model or modify the bidirectional architecture. It only constrains inference to reveal the generated tokens from left to right, token by token. Figure~\ref{fig:ar_over_ao} reports F1 on the Spotify dataset under different bundle sizes. AR consistently outperforms AO across all settings, showing that a fixed prefix-to-suffix order enables the same diffusion recommender to recover more relevant items. These results suggest that arbitrary order decoding can restrict the recommendation space that is effectively explored, despite its seemingly greater generation flexibility. A natural question is: \ul{\textit{why does greater decoding flexibility lead to weaker recommendation quality?}}

\paragraph{Mechanism: Entropy Degradation at Semantic Forks.}

To answer this question, we examine how the four SID tokens are resolved during denoising. The residual quantized tokenizer assigns a hierarchical role to each SID position. As shown in Figure~\ref{fig:dlm_recommender}, the first token selects a coarse semantic region of the item catalog and therefore acts as a \textit{semantic fork} that separates major recommendation branches. The remaining tokens encode residual information that progressively identifies a specific item within the selected region, and we refer to them as \textit{semantic refinement} tokens. Since several coarse branches may remain plausible before the first token is determined, the semantic fork typically has higher entropy than the refinement tokens. As shown in Figure~\ref{fig:method_insight}, confidence-guided AO decoding preferentially reveals low-entropy refinement tokens while postponing the uncertain semantic fork. AR decoding instead resolves the fork before its refinements, forcing the model to confront the major semantic decision early.

Deferring the semantic fork gradually removes the uncertainty required for recommendation exploration. Figure~\ref{fig:sfork_ereduce} reports the mean entropy of each SID position at the step when it is decoded by the same model under AR and AO. Under AR, the first token is resolved before the refinement tokens and retains high entropy, indicating that multiple recommendation branches remain available. Under AO, the refinement tokens are determined first and provide additional context that constrains the possible values of the first token. Consequently, the entropy of the semantic fork decreases substantially when it is eventually decoded, while the refinement tokens do not exhibit the same pronounced reduction. We refer to this position-specific reduction as entropy degradation. \ul{\textit{The model is therefore no longer selecting among several plausible semantic branches, but choosing a fork token that is compatible with the refinement tokens already fixed.}} Across multiple recommendation slots, this process can repeatedly narrow the generated items toward a limited set of branches, producing concentrated recommendations and repeated items as shown in Figure~\ref{fig:motivation_high_sim}.

\subsection{The Quality and Efficiency Tradeoff}

The quality and efficiency tradeoff in diffusion recommendation arises because faster decoding replaces the chain rule factorization of dependent SID tokens with a conditional independence approximation. Let $\mathbf{z}=[z_1,\ldots,z_M]$ denote an SID, $\mathbf{c}$ denote the observed bundle context, and $\pi$ denote a reveal order. When one token is decoded at each step, the resulting sequential policy admits the chain rule factorization

\begin{equation}
\pi_{\theta}^{\pi}
\left[
\mathbf{z}
\mid
\mathbf{c}
\right]
=
\prod_{m=1}^{M}
p_{\theta}
\left[
z_{\pi_m}
\mid
\mathbf{c},
z_{\pi_1},
\ldots,
z_{\pi_{m-1}}
\right].
\end{equation}

so each SID token can condition on all previously resolved tokens. Parallel decoding instead reveals a token set $\mathcal{S}$ from the same visible state and approximates its joint conditional distribution as

\begin{equation}
p_{\theta}
\left[
\mathbf{z}_{\mathcal{S}}
\mid
\mathbf{c},
\mathbf{z}_{\mathcal{O}}
\right]
\approx
\prod_{m\in\mathcal{S}}
p_{\theta}
\left[
z_m
\mid
\mathbf{c},
\mathbf{z}_{\mathcal{O}}
\right],
\end{equation}

where $\mathcal{O}$ contains tokens resolved in earlier rounds. As illustrated in Figure~\ref{fig:motivation_tradeoff}, conservative decoding preserves dependencies across SID levels and progressively constructs a valid catalog identifier, but requires more denoising steps. Aggressive decoding reduces latency by resolving multiple tokens together, yet their individually plausible predictions can form an incompatible SID outside the support of the item catalog.

\begin{figure*}
    \centering
    \includegraphics[width=1\textwidth]{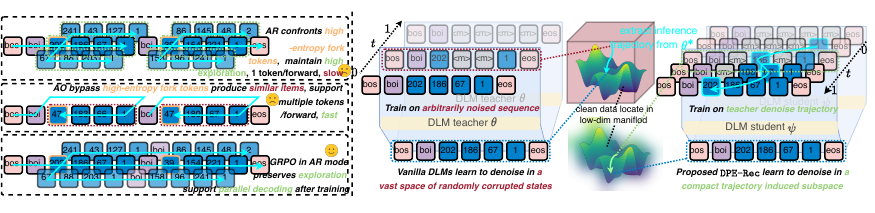}
    \caption{\textit{left:} Overview of autoregressive GRPO for diversity preservation. Left to right rollouts resolve semantic forks before refinement without altering the bidirectional architecture; \textit{right:} Overview of pseudo trajectory distillation. A sequential AO teacher provides high quality denoising trajectories, and the student learns their induced states to decode multiple confident tokens in parallel. 
   }
    \label{fig:method_insight}
    \vspace{-10pt}
\end{figure*}

This tradeoff is rooted in the standard random masking objective, which trains the denoiser to reconstruct clean sequences from a vast combinatorial space of corrupted states. Each mask pattern defines a different conditional prediction problem, and the number of such patterns grows exponentially with sequence length. However, as shown in Figure~\ref{fig:method_insight}, a fixed inference policy visits only a compact trajectory-induced subspace along its denoising trajectories. Covering the much broader corruption space, therefore, imposes a largely unnecessary learning burden and disperses denoising capacity across states that are rarely or never encountered during inference. This diffuse supervision leaves the model insufficiently prepared to preserve token dependencies along the short trajectories required for efficient decoding.
\section{Method}
To address the limitations analyzed above, \name preserves diversity by confronting semantic forks through autoregressive GRPO and improves efficiency by concentrating denoising learning from the vast random corruption space onto the compact trajectory induced subspace through pseudo trajectory distillation.

\subsection{Confronting Semantic Forks: Autoregressive GRPO for Exploration Preservation}

Confidence-guided decoding restricts recommendation exploration by postponing semantic forks until their possible values have already been constrained by semantic refinement. To confront semantic forks while their uncertainty remains intact, \name induces an autoregressive SID policy from the pretrained diffusion denoiser and optimizes this policy with GRPO. The autoregressive order is used only as a post-training scaffold, leaving the bidirectional architecture and its parallel generation capability unchanged.

Let the target segment contain $N$ item slots, with each item represented by an SID of $M$ tokens. We serialize the target segment as an output sequence
$o=[o_1,\ldots,o_{MN}]$ using a fixed order that resolves the semantic fork before the refinement tokens within each item slot. 
Let $m(k)\in\{1,\ldots,M\}$ denote the index of token $o_k$ within its corresponding SID, which we refer to as its semantic level. At the $k$th decision, the model input $\mathbf{x}^{(k)}$ contains the observed bundle $\mathbf{c}$, the special tokens marking sequence and item boundaries, and the previously generated tokens $o_{<k}$, while token $o_k$ and all subsequent target tokens remain masked. The denoiser distribution at position $k$ therefore defines the autoregressive policy

\begin{equation}
\pi_{\theta}
\left(
o_k=v
\mid
o_{<k},\mathbf{c}
\right)
=
\frac{
\exp
\left(
s_{\theta,k}
\left(
v\mid\mathbf{x}^{(k)}
\right)
/\tau_{m(k)}
\right)
}{
\sum_{v'\in\mathcal{V}_{m(k)}}
\exp
\left(
s_{\theta,k}
\left(
v'\mid\mathbf{x}^{(k)}
\right)
/\tau_{m(k)}
\right)
},
\end{equation}

where $s_{\theta,k}$ denotes the denoiser logit, $\mathcal{V}_{m(k)}$ denotes the valid vocabulary of SID level $m(k)$, and $\tau_{m(k)}$ controls its sampling temperature. We assign greater stochasticity to the semantic fork and use more conservative sampling for semantic refinement. This asymmetric policy concentrates exploration on decisions that select recommendation branches rather than indiscriminately perturbing all SID levels.

The induced policy admits the exact autoregressive factorization

\begin{equation}
\pi_{\theta}
\left(
o\mid\mathbf{c}
\right)
=
\prod_{k=1}^{MN}
\pi_{\theta}
\left(
o_k\mid o_{<k},\mathbf{c}
\right).
\end{equation}

This factorization provides tractable token likelihoods for GRPO without introducing a causal model or modifying the attention structure. Resolving each semantic fork before its refinement tokens preserves uncertainty at the branch decision and exposes the policy to alternative recommendation paths.

To evaluate each rollout, we define a hierarchical set reward that measures recommendation agreement at progressively finer semantic levels. For rollout $o_i$, let

\begin{equation}
\widehat{\mathcal{Z}}_{i}^{\ell}
=
\left\{
\hat{\mathbf{z}}_{i,1:\ell}^{(n)}
\right\}_{n=1}^{N},
\qquad
\mathcal{Z}_{\star}^{\ell}
=
\left\{
\mathbf{z}_{1:\ell}^{\star(n)}
\right\}_{n=1}^{N}
\end{equation}

denote the sets of generated and ground truth SIDs truncated after level $\ell$. The reward is defined as

\begin{equation}
\begin{aligned}
r(o_i)
={}
w_1
\frac{
\left|
\widehat{\mathcal{Z}}_{i}^{2}
\cap
\mathcal{Z}_{\star}^{2}
\right|
}{
N
}
+
w_2
\frac{
\left|
\widehat{\mathcal{Z}}_{i}^{M}
\cap
\mathcal{Z}_{\star}^{M}
\right|
}{
N
}.
\end{aligned}
\end{equation}

The two terms reward finer semantic agreement, and exact item recovery. Since each term is computed over sets, repeated predictions receive no additional credit, while the coarser SID levels provide informative rewards when exact item matches remain sparse.

For each bundle context, we sample a group of rollouts from the old autoregressive policy and optimize them using the GRPO objective introduced in Section~\ref{sec:prelim_grpo}. The resulting training process confronts semantic forks before entropy degradation can suppress alternative branches and rewards exploration that recovers diverse and relevant items. Since autoregressive generation is used only to define the post-training policy, the optimized denoiser retains its original bidirectional structure for subsequent parallel decoding.

\subsection{Breaking the Pareto Front via Trajectory Distillation}

The quality and efficiency frontier arises because standard random masking distributes denoising supervision across a vast combinatorial space of corrupted states, while efficient decoding requires reliable predictions within a compact trajectory-induced subspace. To concentrate denoising capacity on this subspace, \name distills high-quality, arbitrary-order trajectories from the GRPO-optimized denoiser into a student model. The teacher $\theta$ resolves one token per step to preserve token dependencies, while the student $\psi$ learns to reproduce multiple future decisions from intermediate states along the same trajectory.

\paragraph{High Quality Teacher Trajectory Extraction.}

We use the GRPO-optimized denoiser as the teacher and decode each target segment with confidence-guided arbitrary-order decoding. Unlike the autoregressive order used for policy optimization, the teacher dynamically selects the most confident unresolved position at every step. Let $\mathcal{M}^{(j)}$ denote the masked positions in the teacher state $\mathbf{x}^{(j)}$. The confidence of position $k$ is

\begin{equation}
\kappa_{k}^{(j)}
=
\max_{v\in\mathcal{V}_{m(k)}}
p_{\theta}
\left(
v
\mid
\mathbf{x}^{(j)}
\right),
\end{equation}

and the next position is selected as

\begin{equation}
\pi_j
=
\arg\max_{k\in\mathcal{M}^{(j)}}
\kappa_k^{(j)}.
\end{equation}

Only position $\pi_j$ is resolved before the next denoising step. The teacher therefore retains the flexibility of arbitrary order decoding while allowing every decision to condition on all tokens committed in earlier rounds. This sequential execution avoids the conditional independence approximation among simultaneously revealed tokens and produces a high-quality completion $\tilde{\mathbf{x}}_0$ together with its reveal order
$\pi=[\pi_1,\ldots,\pi_{MN}]$.

\paragraph{Trajectory Induced State Construction.}

The teacher trajectory identifies the intermediate states that arise under high-quality diffusion decoding. For a masking ratio $t$, we retain the first

\begin{equation}
r(t)
=
\left\lfloor
(1-t)MN
\right\rfloor
\end{equation}

teacher decisions and mask the remaining target positions. The resulting state is

\begin{equation}
\tilde{x}_{t,k}
=
\begin{cases}
\tilde{x}_{0,k},
&
k\in
\{\pi_1,\ldots,\pi_{r(t)}\},
\\
[\mathrm{MASK}],
&
k\in
\{\pi_{r(t)+1},\ldots,\pi_{MN}\}.
\end{cases}
\end{equation}

The observed bundle context and structural tokens remain unchanged. Each
$\tilde{\mathbf{x}}_t$ therefore lies on the trajectory induced by the teacher decoder, rather than representing an arbitrary corruption pattern sampled from the much broader random masking space.

We progressively increase the difficulty of these states during training. Let $u\in[0,1]$ denote the training progress. We sample

\begin{equation}
t
\sim
\mathcal{U}
\left[
t_{\min}
+
u
\left(
t_{\max}-t_{\min}
\right),
t_{\max}
\right].
\end{equation}

Early training emphasizes later trajectory states with fewer unresolved tokens, while subsequent training gradually exposes the student to earlier states that require longer-range denoising. We also construct a complementary view by masking the first $r(t)$ teacher decisions and revealing the remaining positions, providing supervision across both portions of the same trajectory partition.

\paragraph{Trajectory Distillation Objective.}

The student is initialized from the teacher and trained to recover the teacher completion from trajectory-induced states. Let $\mathcal{M}_t$ denote the masked positions in $\tilde{\mathbf{x}}_t$. The primary objective is

\begin{equation}
\mathcal{L}_{\mathrm{CE}}
=
-
\frac{1}{|\mathcal{M}_t|}
\sum_{k\in\mathcal{M}_t}
\log
p_{\psi}
\left(
\tilde{x}_{0,k}
\mid
\tilde{\mathbf{x}}_t
\right).
\end{equation}

We define $\mathcal{L}_{\mathrm{CE}}^{\mathrm{comp}}$ analogously for the complementary view. These two objectives train the student to recover multiple future teacher decisions from different states along the same denoising trajectory.

The complete objective is

\begin{equation}
\mathcal{L}_{\mathrm{TD}}
=
\mathcal{L}_{\mathrm{CE}}
+
\mathcal{L}_{\mathrm{CE}}^{\mathrm{comp}}
.
\end{equation}

\paragraph{Confidence Guided Parallel Inference.}

At inference, the distilled student commits all masked positions whose confidence exceeds a threshold. Let $\mathcal{M}_s$ denote the masked positions at round $s$. The set of positions revealed in parallel is

\begin{equation}
\mathcal{S}_s
=
\left\{
k\in\mathcal{M}_s
\mid
\max_{v\in\mathcal{V}_{m(k)}}
p_{\psi}
\left(
v
\mid
\mathbf{x}^{(s)}
\right)
\ge
\eta
\right\}
\cup
\operatorname{Top}_{q_s}
\left(
\mathcal{M}_s
\right),
\end{equation}

\begin{equation}
q_s
=
\left\lceil
\frac{
|\mathcal{M}_s|
}{
F-s
}
\right\rceil,
\end{equation}

where $F$ denotes the denoising step budget and $q_s$ guarantees that all remaining positions are resolved within this budget. The decoder preserves the arbitrary order behavior of the teacher but resolves multiple sufficiently reliable positions in each round.

Trajectory distillation thus transfers the token dependencies resolved by high-quality sequential AO decoding into a compact number of parallel denoising transitions. By redistributing denoising capacity from the vast random corruption space to the trajectory-induced subspace encountered during inference, \name preserves recommendation quality with substantially fewer denoising steps.

\subsection{Item Legalization}
\label{sec:item_legalization}

Following the SID-to-item mapping strategy of DDBC~\cite{tu2026discrete}, we apply item legalization  within the candidate set discussed in Sec~\ref{sec:evaluation_protocal} after generation. A generated SID is mapped to its exact catalog item when available; otherwise, its representation is reconstructed from the semantic codebooks and matched to the candidate item with the highest cosine similarity. Selected candidates are removed to ensure distinct recommendations.
\section{Experiment}

We conduct comprehensive experiments to assess the effectiveness of \name in diffusion-based bundle construction. Our evaluation is organized around the following research questions:

\textbf{RQ1: Overall Performance.} How does \name compare with existing scoring-based, autoregressive, and diffusion-based methods in terms of recommendation quality and inference efficiency?

\textbf{RQ2: Where Should Exploration Be Applied in Hierarchical SIDs?}
Does concentrating sampling uncertainty on the semantic fork improve recommendation quality and bundle diversity?

\textbf{RQ3: Breaking the Quality and Efficiency Tradeoff.}
Can pseudo trajectory distillation reduce denoising steps without sacrificing SID validity?

\textbf{RQ4: Ablation Study.} How does each component of \name contribute to its recommendation quality, diversity preservation, and inference efficiency?

\subsection{Experimental Setup}
\label{sec:experimental_setup}
\paragraph{Datasets.}
We evaluate \name on Spotify \cite{chen2018recsys} for playlist continuation and on POG \cite{chen2019pog} and IQON3000 \cite{song2019gpbpr} for outfit completion. We use five evaluation settings: Spotify$_{30}$, Spotify$_{60}$, Spotify$_{90}$, four-item outfits from POG, and eight-item outfits from IQON3000. Each evaluation bundle is split evenly into an observed subset and a target subset for completion, with disjoint training, validation, and test bundles.
All content-aware methods share the same fixed 64-dimensional CLHE \cite{ma2024leveraging} item representations, and all methods with fixed SIDs share the same four-token SIDs. Dataset statistics, preprocessing, augmentation, and tokenization details are provided in Appendix~\ref{sec:dataset_detail}.
\paragraph{Baselines.}
We compare \name with representative baselines from three families. \textbf{ID based sequential models} include GRU4Rec\cite{hidasi2015session}, SASRec\cite{kang2018self}, HSTU\cite{zhai2024actions}, and BERT4Rec\cite{sun2019bert4rec}. \textbf{Representation enhanced discriminative models} include FDSA\cite{zhang2019feature} and RecJPQ\cite{petrov2024recjpq}, which enrich item modeling with feature information or compositional representations. \textbf{Semantic ID based generative models} include DiffGRM\cite{liu2026diffgrm}, UniGRec\cite{li2026unigrec}, and DDBC\cite{tu2026discrete}, which generate target items through discrete item identifiers. For training details, please see Appendix~\ref{sec:training_detail}

\paragraph{Evaluation Metrics.}
We evaluate recommendation quality using F1, Jaccard similarity, and optimal assignment similarity (OAS). The first two measure exact set agreement, while OAS computes the mean cosine similarity under optimal one-to-one matching between predicted and target items. All methods are trained and evaluated using five random seeds, with results reported as the mean and standard deviation. Detailed metric definitions are provided in Appendix~\ref{sec:metric_detail}.

\paragraph{Evaluation Protocol.}
\label{sec:evaluation_protocal}
Following sampled-candidate evaluation in prior recommendation studies~\cite{he2017neural,kang2018self,sun2019bert4rec} and the bundle construction protocol of DDBC~\cite{tu2026discrete}, we evaluate all methods on shared candidate sets. With large catalogs and few ground-truth targets per bundle, full-catalog exact-match scores can be very low. For each test bundle, we construct a candidate set containing all target items and uniformly sampled non-target items, with a total size equal to one hundred times the target bundle size. F1, Jaccard similarity, and OAS are then computed.

\subsection{Main Results}
\begin{table*}[htbp]
\centering
\caption{Recommendation quality on Spotify, POG, and IQON3000 (mean $\pm$ standard deviation over five random seeds). The best and second best results are highlighted in \textbf{bold} and \ul{underlined}, respectively. Average improvement is computed relative to all baselines.}
\label{tab:main_results}
\renewcommand{\arraystretch}{1}
\setlength{\tabcolsep}{3.5pt}
\providecommand{\ms}{}
\renewcommand{\ms}[2]{\vtop{\setbox0\hbox{#1}\dp0=1.15ex\copy0\hbox to\wd0{\hss{\footnotesize$\pm$#2}\hss}\kern0.85ex}}
\resizebox{\linewidth}{!}{%
\begin{tabular}{c*{5}{>{\hspace{7pt}}ccc}}
\toprule
 & \multicolumn{3}{>{\hspace{7pt}}c}{Spotify K=30} & \multicolumn{3}{>{\hspace{7pt}}c}{Spotify K=60} & \multicolumn{3}{>{\hspace{7pt}}c}{Spotify K=90} & \multicolumn{3}{>{\hspace{7pt}}c}{POG K=4} & \multicolumn{3}{>{\hspace{7pt}}c}{IQON3000 K=8} \\
\cmidrule(l{10.5pt}r{3.5pt}){2-4}\cmidrule(l{10.5pt}r{3.5pt}){5-7}\cmidrule(l{10.5pt}r{3.5pt}){8-10}\cmidrule(l{10.5pt}r{3.5pt}){11-13}\cmidrule(l{10.5pt}r{3.5pt}){14-16}
model & F1 ↑ & Jaccard ↑ & OAS ↑ & F1 ↑ & Jaccard ↑ & OAS ↑ & F1 ↑ & Jaccard ↑ & OAS ↑ & F1 ↑ & Jaccard ↑ & OAS ↑ & F1 ↑ & Jaccard ↑ & OAS ↑ \\ \midrule
GRU4Rec & \ms{0.226}{0.003} & \ms{0.143}{0.002} & \ms{0.532}{0.003} & \ms{0.168}{0.002} & \ms{0.101}{0.001} & \ms{0.510}{0.001} & \ms{0.140}{0.006} & \ms{0.081}{0.004} & \ms{0.495}{0.007} & \ms{0.144}{0.003} & \ms{0.101}{0.002} & \ms{0.515}{0.006} & \ms{0.029}{0.001} & \ms{0.017}{0.001} & \ms{0.200}{0.001} \\
SASRec & \ms{0.168}{0.008} & \ms{0.102}{0.006} & \ms{0.464}{0.009} & \ms{0.107}{0.004} & \ms{0.060}{0.002} & \ms{0.429}{0.004} & \ms{0.113}{0.008} & \ms{0.063}{0.005} & \ms{0.449}{0.009} & \ms{0.100}{0.033} & \ms{0.067}{0.022} & \ms{0.384}{0.021} & \ms{0.045}{0.003} & \ms{0.026}{0.002} & \ms{0.206}{0.008} \\
BERT4Rec & \ms{0.238}{0.011} & \ms{0.152}{0.008} & \ms{0.523}{0.009} & \ms{0.248}{0.014} & \ms{0.156}{0.010} & \ms{0.563}{0.014} & \ms{0.217}{0.064} & \ms{0.135}{0.044} & \ms{0.544}{0.065} & \ms{0.178}{0.003} & \ms{0.122}{0.002} & \ms{0.493}{0.006} & \ms{0.064}{0.002} & \ms{0.038}{0.001} & \ms{0.207}{0.004} \\
FDSA & \ms{0.347}{0.006} & \ms{0.230}{0.005} & \ms{0.641}{0.005} & \ms{0.310}{0.007} & \ms{0.198}{0.005} & \ms{0.636}{0.006} & \ms{0.290}{0.011} & \ms{0.183}{0.008} & \ms{0.634}{0.009} & \ms{0.177}{0.004} & \ms{0.126}{0.002} & \ms{0.531}{0.007} & \ms{0.051}{0.006} & \ms{0.031}{0.004} & \ms{0.217}{0.014} \\
RecJPQ & \ms{0.142}{0.003} & \ms{0.084}{0.002} & \ms{0.446}{0.003} & \ms{0.155}{0.005} & \ms{0.092}{0.003} & \ms{0.512}{0.005} & \ms{0.169}{0.027} & \ms{0.103}{0.018} & \ms{0.522}{0.013} & \ms{0.122}{0.014} & \ms{0.083}{0.009} & \ms{0.475}{0.007} & \ms{0.039}{0.001} & \ms{0.023}{0.001} & \ms{0.194}{0.004} \\
HSTU & \ms{0.254}{0.010} & \ms{0.163}{0.007} & \ms{0.549}{0.010} & \ms{0.216}{0.010} & \ms{0.135}{0.006} & \ms{0.544}{0.010} & \ms{0.190}{0.014} & \ms{0.116}{0.009} & \ms{0.536}{0.012} & \ms{\textbf{0.202}}{0.004} & \ms{\textbf{0.146}}{0.003} & \ms{0.523}{0.008} & \ms{0.076}{0.001} & \ms{0.046}{0.002} & \ms{0.252}{0.010} \\
UnigRec & \ms{0.360}{0.002} & \ms{0.238}{0.002} & \ms{0.645}{0.002} & \ms{\ul{\textit{0.369}}}{0.008} & \ms{\ul{\textit{0.241}}}{0.006} & \ms{\ul{\textit{0.678}}}{0.005} & \ms{0.361}{0.003} & \ms{0.234}{0.002} & \ms{0.686}{0.004} & \ms{\ul{\textit{0.200}}}{0.004} & \ms{\ul{\textit{0.142}}}{0.004} & \ms{0.529}{0.022} & \ms{0.126}{0.006} & \ms{0.079}{0.004} & \ms{0.323}{0.005} \\
DiffGRM & \ms{\ul{\textit{0.371}}}{0.006} & \ms{\textbf{0.248}}{0.004} & \ms{\ul{\textit{0.652}}}{0.004} & \ms{0.361}{0.004} & \ms{0.236}{0.003} & \ms{0.670}{0.005} & \ms{\ul{\textit{0.368}}}{0.006} & \ms{\ul{\textit{0.240}}}{0.004} & \ms{\ul{\textit{0.692}}}{0.007} & \ms{0.192}{0.006} & \ms{0.139}{0.005} & \ms{\textbf{0.537}}{0.008} & \ms{0.128}{0.004} & \ms{0.081}{0.002} & \ms{0.316}{0.008} \\
DDBC & \ms{0.277}{0.002} & \ms{0.174}{0.002} & \ms{0.615}{0.001} & \ms{0.284}{0.002} & \ms{0.176}{0.001} & \ms{0.661}{0.001} & \ms{0.280}{0.001} & \ms{0.172}{0.000} & \ms{0.681}{0.001} & \ms{0.138}{0.002} & \ms{0.098}{0.002} & \ms{0.526}{0.002} & \ms{0.054}{0.001} & \ms{0.032}{0.001} & \ms{0.268}{0.001} \\ \midrule
\rowcolor[HTML]{C0C0C0}
\name & \ms{\textbf{0.372}}{0.003} & \ms{\ul{\textit{0.247}}}{0.002} & \ms{\textbf{0.659}}{0.002} & \ms{\textbf{0.378}}{0.002} & \ms{\textbf{0.248}}{0.002} & \ms{\textbf{0.697}}{0.002} & \ms{\textbf{0.379}}{0.001} & \ms{\textbf{0.246}}{0.001} & \ms{\textbf{0.713}}{0.001} & \ms{0.198}{0.006} & \ms{\ul{\textit{0.142}}}{0.005} & \ms{0.533}{0.004} & \ms{\ul{\textit{0.131}}}{0.001} & \ms{\ul{\textit{0.082}}}{0.001} & \ms{\ul{\textit{0.331}}}{0.002} \\
Avg. Improv. (\%) & 54.8\% & 64.0\% & 19.3\% & 79.1\% & 93.7\% & 23.2\% & 86.4\% & 101.4\% & 25.3\% & 29.5\% & 33.0\% & 7.4\% & 139.6\% & 150.6\% & 41.2\% \\
\rowcolor[HTML]{C0C0C0}
\name-Fast & \ms{0.360}{0.004} & \ms{0.239}{0.003} & \ms{0.635}{0.004} & \ms{0.364}{0.003} & \ms{0.238}{0.002} & \ms{0.665}{0.002} & \ms{0.358}{0.002} & \ms{0.231}{0.002} & \ms{0.678}{0.002} & \ms{\ul{\textit{0.200}}}{0.004} & \ms{\ul{\textit{0.142}}}{0.003} & \ms{\ul{\textit{0.536}}}{0.002} & \ms{\textbf{0.146}}{0.002} & \ms{\textbf{0.092}}{0.001} & \ms{\textbf{0.342}}{0.004} \\
Avg. Improv. (\%) & 49.9\% & 58.4\% & 14.8\% & 72.3\% & 85.9\% & 17.6\% & 76.1\% & 89.2\% & 19.2\% & 31.0\% & 33.3\% & 7.9\% & 167.6\% & 183.2\% & 46.0\% \\ \bottomrule
\end{tabular}%
}
\end{table*}

\begin{table}[htbp]
\centering
\caption{Inference latency per bundle on Spotify, POG, and IQON3000. Lower values indicate faster inference. Average improvement is computed relative to the mean latency of all baselines.}
\label{tab:speed_up}
\renewcommand{\arraystretch}{0.85}
\resizebox{\columnwidth}{!}{%
\begin{tabular}{@{}cccccc@{}}
\toprule
\multicolumn{6}{c}{Latency   (ms) ↓}                                                                                 \\ \midrule
model                     & Spotify              & Spotify              & Spotify              & POG                 & IQON3000      \\
                          & {\small K=30}        & {\small K=60}        & {\small K=90}        & {\small K=4}        & {\small K=8}              \\ \midrule
GRU4Rec                   & \ul{ \textit{13.18}} & \ul{ \textit{26.17}} & \ul{ \textit{39.10}} & \ul{ \textit{1.53}} & 5.92                      \\
SASRec                    & 26.82                & 55.84                & 82.95                & 3.90                & 7.71                      \\
BERT4Rec                  & \textbf{1.66}        & \textbf{1.93}        & \textbf{2.27}        & \textbf{1.39}       & \textbf{1.51}             \\
FDSA                      & 37.79                & 75.53                & 113.41               & 5.24                & 9.77                      \\
RecJPQ                    & 22.15                & 44.03                & 66.01                & 3.16                & 6.01                      \\
HSTU                      & 18.54                & 36.16                & 54.91                & 2.72                & \ul{\textit{5.26}}        \\
UniGRec                   & 593.76               & 1168.84              & 1792.86              & 80.35               & 154.68                    \\
DiffGRM                   & 367.89               & 736.45               & 1267.27              & 68.47               & 91.40                     \\
DDBC                      & 2930.87              & 5822.53              & 8262.87              & 549.53              & 872.57                    \\ \midrule
\rowcolor[HTML]{C0C0C0} 
\name      & 443.59               & 903.41               & 1340.05              & 59.38               & 112.51                    \\
Avg. Speedup ($\times$)
& 1.01 & 0.98 & 0.97 & 1.34 & 1.14 \\
\rowcolor[HTML]{C0C0C0} 
\name-Fast & 62.34                & 65.79                & 66.72                & 46.49               & 58.32                     \\
Avg. Speedup ($\times$)
& 7.15 & 13.46 & 19.45 & 1.71 & 2.20 \\ \bottomrule
\end{tabular}%
}
\end{table}

Table~\ref{tab:main_results} shows that \name and \name-Fast achieve approximately 63\% average relative improvement over the nine baselines across the five evaluation settings and three metrics. We attribute these gains to two factors. First, the DLM's bidirectional attention provides an inductive bias better aligned with unordered bundle construction. Second, our autoregressive GRPO preserves exploration by confronting semantic forks before refinement tokens constrain their choices, allowing the model to explore alternative recommendation branches.

Table~\ref{tab:speed_up} shows that \name-Fast achieves an average speedup of 25.6$\times$ over the three generative baselines across the five evaluation settings, while retaining competitive recommendation quality. Its latency increases only from 62.34 to 66.72 ms as the Spotify bundle size grows from 30 to 90. Pseudo-trajectory distillation concentrates learning on teacher-visited denoising states, improving the reliability of multi-token predictions. Confidence-thresholded decoding then exploits this capability to resolve multiple tokens in parallel, substantially reducing sequential denoising rounds. The additional training cost of GRPO and PTD and its amortization through inference savings are analyzed in Appendix~\ref{sec:training_cost}.

\subsection{Where Should Exploration Be Applied in Hierarchical SIDs?}
\begin{figure}
    \centering
    \includegraphics[width=1\columnwidth]{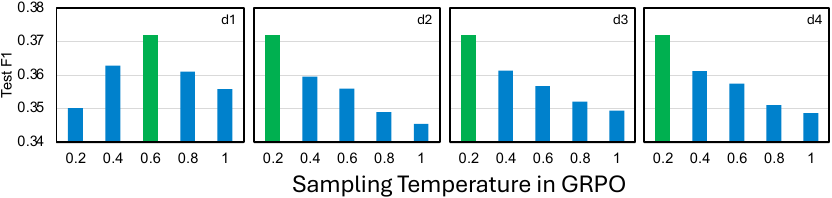}
    \caption{Effect of level specific sampling temperature during autoregressive GRPO. Test F1 peaks at $0.6$ for the semantic fork $d_1$, while the refinement levels $d_2$ to $d_4$ perform best at $0.2$.
   }
    \label{fig:temp_sweep}
\end{figure}
\begin{figure}
    \centering
    \vspace{-10pt}
    \includegraphics[width=1\columnwidth]{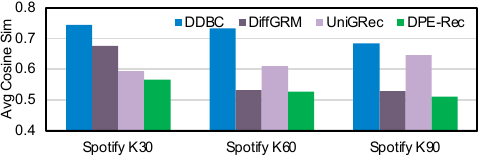}
    \caption{Mean pairwise cosine similarity among generated items across different Spotify bundle sizes. \name achieves the lowest similarity among the compared methods across all three settings, indicating greater bundle diversity and aligning with its higher F1 performance.
   }
    \label{fig:cross_item_similarity}
    \vspace{-10pt}
\end{figure}
We investigate how sampling temperature should be distributed across the four SID positions during autoregressive GRPO. We vary the temperature of one position at a time while keeping the remaining training settings fixed. As shown in Figure~\ref{fig:temp_sweep}, the first SID position achieves the highest F1 at a moderate temperature of $0.6$. In contrast, the remaining positions perform best at the lowest tested temperature of $0.2$, and their performance consistently decreases as the temperature increases. Consistent with Section~\ref{sec:flexibility_trap}, we therefore set $\tau_1=0.6$ and $\tau_2=\tau_3=\tau_4=0.2$ to concentrate exploration at the semantic fork while preserving item consistency at later positions.

Figure~\ref{fig:cross_item_similarity} shows that \name consistently produces lower pairwise cosine similarity than DDBC, DiffGRM, and UniGRec across all three Spotify settings. Together with the temperature sweep, these results support concentrating exploration at semantic forks, where alternative recommendation branches remain available before refinement tokens constrain their choices. Additional diversity results, including catalog coverage, and cluster entropy, are reported in Appendix~\ref{sec:diversity_analysis}.
\vspace{-5pt}
\subsection{Breaking the Quality and Efficiency Tradeoff}
\begin{figure}
    \centering
    \includegraphics[width=1\columnwidth]{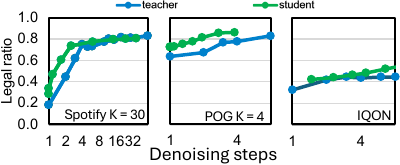}
    \vspace{-15pt}
    \caption{Legal SID ratio under different denoising budgets on Spotify$_{30}$, POG and IQON3000. The distilled student maintains higher validity than the teacher at comparable numbers of denoising steps, demonstrating an improved quality and efficiency tradeoff.
   }
    \label{fig:distillation_break_tradeoff}
    \vspace{-15pt}
\end{figure}
Figure~\ref{fig:distillation_break_tradeoff} compares the teacher and distilled student under different denoising budgets on Spotify$_{30}$, POG, and IQON3000. Aggressive parallel decoding sharply reduces the teacher's legal ratio, whereas the student maintains a substantially higher legal ratio at comparable numbers of denoising steps. By concentrating learning on teacher-visited denoising states, PTD improves the reliability of multi-token predictions, enabling faster decoding while better preserving SID validity.

\subsection{Ablation Study}
\vspace{-5pt}
\begin{table}[htbp]
\centering
\caption{Ablation study averaged over Spotify$_{30}$, Spotify$_{60}$, Spotify$_{90}$, POG, and IQON3000. PTD refers to pseudo trajectory distillation.}
\label{tab:ablation}
\renewcommand{\arraystretch}{0.85}
\resizebox{\columnwidth}{!}{%
\begin{tabular}{@{}ccccc@{}}
\toprule
model           & F1 ↑     & Jaccard ↑ & OAS ↑    & Latency (ms) ↓ \\ \midrule
Vanilla DLM Rec & 0.226    & 0.145     & 0.558    & 571.79         \\
+GRPO           & +28.86\% & +33.29\%  & +5.06\%  & +0.00\%        \\
+PTD            & -2.02\%  & -2.32\%   & -2.63\%  & -89.52\%       \\ \bottomrule
\end{tabular}%
}
\end{table}

Table~\ref{tab:ablation} reports cumulative ablations averaged over the five evaluation settings. Adding fork-aware GRPO improves F1 and Jaccard similarity by 28.86\% and 33.29\%, respectively, without increasing inference latency. PTD further reduces latency by 89.52\%, with only about 2.5\% decreases across the three quality metrics. These results demonstrate that trajectory-focused distillation translates the quality benefits of fork-aware exploration into fast parallel decoding, retaining strong recommendation quality while cutting inference latency by nearly an order of magnitude. Fine-grained reward and group-size analyses are provided in Appendix~\ref{sec:detail_ablation}.
\section{Related Work}


\subsection{Bundle Construction}

Bundle construction selects items from large catalogs to generate or complete bundles. BGGN models user, item, and bundle relations with graph neural networks for personalized bundle generation, while BundleMage learns bundle completion jointly with bundle matching \cite{chang2021bundle,jeon2023accurate}. Bundle MLLM incorporates multimodal item information through a fine tuned multimodal language model \cite{liu2025fine}. DDBC enables nonsequential bundle construction through masked discrete diffusion over semantic identifiers \cite{tu2026discrete}. Our work studies the diversity and efficiency limitations introduced by the denoising process and addresses them through autoregressive GRPO and pseudo trajectory distillation.
\subsection{Generative Recommendation}

Recommendation methods differ in item representation and target prediction. ID based sequential models predict item identities from interaction histories. GRU4Rec\cite{hidasi2015session} models item transitions with recurrent networks, SASRec\cite{kang2018self} captures sequential dependencies through self attention, HSTU\cite{zhai2024actions} targets large scale sequential recommendation, and BERT4Rec\cite{sun2019bert4rec} learns bidirectional representations through masked item prediction. Representation enhanced discriminative models enrich item representations: FDSA\cite{zhang2019feature} models item and feature transitions, while RecJPQ\cite{petrov2024recjpq} uses shared sub item embeddings. Semantic ID based generative models encode items as discrete token sequences. UniGRec\cite{li2026unigrec} jointly optimizes soft identifiers and recommendation, DiffGRM\cite{liu2026diffgrm} applies masked diffusion to next item generation, and DDBC\cite{tu2026discrete} extends it to bundle construction. In contrast, \name focuses on diversity preservation and efficient inference when jointly generating multiple bundle items.

\subsection{Diffusion Language Models}

Diffusion language models generate discrete sequences through iterative masked denoising. MDLM\cite{sahoo2024simple} establishes a simple masked diffusion language modeling framework, while LLaDA\cite{nie2026large} scales this paradigm through pretraining and instruction tuning. Recent studies\cite{arriola2025block,wu2025fast,hu2025flashdlm} accelerate inference through block diffusion, parallel decoding, and cache reuse. CD4LM\cite{liang2026cd4lm} reduces denoising steps through consistency distillation and adaptive decoding. Trajectory distillation further reduces denoising steps \cite{qian2026d3llm,zhang2026few}. Beyond inference efficiency, \citet{ni2026flexibility} identify the flexibility trap in DLM reasoning, where flexible decoding may postpone difficult decisions, and apply GRPO-based post-training to improve reasoning performance.
\section{Conclusion}

We presented \name, a diffusion framework for accurate and efficient bundle construction. Combined with autoregressive GRPO and pseudo trajectory distillation, \name preserves bundle diversity while enabling reliable multi-token parallel decoding. Across playlist and outfit benchmarks, \name improves recommendation performance by 63\% on average, while its fast variant achieves 25.6$\times$ faster inference on average than existing generative recommenders, with minimal loss in recommendation quality.

\bibliographystyle{ACM-Reference-Format}
\bibliography{refs/ref}


\begin{thebibliography}{41}


\ifx \showCODEN    \undefined \def \showCODEN     #1{\unskip}     \fi
\ifx \showISBNx    \undefined \def \showISBNx     #1{\unskip}     \fi
\ifx \showISBNxiii \undefined \def \showISBNxiii  #1{\unskip}     \fi
\ifx \showISSN     \undefined \def \showISSN      #1{\unskip}     \fi
\ifx \showLCCN     \undefined \def \showLCCN      #1{\unskip}     \fi
\ifx \shownote     \undefined \def \shownote      #1{#1}          \fi
\ifx \showarticletitle \undefined \def \showarticletitle #1{#1}   \fi
\ifx \showURL      \undefined \def \showURL       {\relax}        \fi
\providecommand\bibfield[2]{#2}
\providecommand\bibinfo[2]{#2}
\providecommand\natexlab[1]{#1}
\providecommand\showeprint[2][]{arXiv:#2}

\bibitem[Arriola et~al\mbox{.}(2025)]%
        {arriola2025block}
\bibfield{author}{\bibinfo{person}{Marianne Arriola}, \bibinfo{person}{Aaron Gokaslan}, \bibinfo{person}{Justin Chiu}, \bibinfo{person}{Zhihan Yang}, \bibinfo{person}{Zhixuan Qi}, \bibinfo{person}{Jiaqi Han}, \bibinfo{person}{Subham Sahoo}, {and} \bibinfo{person}{Volodymyr Kuleshov}.} \bibinfo{year}{2025}\natexlab{}.
\newblock \showarticletitle{Block diffusion: Interpolating between autoregressive and diffusion language models}. In \bibinfo{booktitle}{\emph{International Conference on Learning Representations}}, Vol.~\bibinfo{volume}{2025}. \bibinfo{pages}{50726--50753}.
\newblock


\bibitem[Chang et~al\mbox{.}(2021)]%
        {chang2021bundle}
\bibfield{author}{\bibinfo{person}{Jianxin Chang}, \bibinfo{person}{Chen Gao}, \bibinfo{person}{Xiangnan He}, \bibinfo{person}{Depeng Jin}, {and} \bibinfo{person}{Yong Li}.} \bibinfo{year}{2021}\natexlab{}.
\newblock \showarticletitle{Bundle recommendation and generation with graph neural networks}.
\newblock \bibinfo{journal}{\emph{IEEE Transactions on Knowledge and Data Engineering}} \bibinfo{volume}{35}, \bibinfo{number}{3} (\bibinfo{year}{2021}), \bibinfo{pages}{2326--2340}.
\newblock


\bibitem[Chen et~al\mbox{.}(2018)]%
        {chen2018recsys}
\bibfield{author}{\bibinfo{person}{Ching-Wei Chen}, \bibinfo{person}{Paul Lamere}, \bibinfo{person}{Markus Schedl}, {and} \bibinfo{person}{Hamed Zamani}.} \bibinfo{year}{2018}\natexlab{}.
\newblock \showarticletitle{Recsys challenge 2018: Automatic music playlist continuation}. In \bibinfo{booktitle}{\emph{Proceedings of the 12th ACM Conference on Recommender Systems}}. \bibinfo{pages}{527--528}.
\newblock


\bibitem[Chen et~al\mbox{.}(2019)]%
        {chen2019pog}
\bibfield{author}{\bibinfo{person}{Wen Chen}, \bibinfo{person}{Pipei Huang}, \bibinfo{person}{Jiaming Xu}, \bibinfo{person}{Xin Guo}, \bibinfo{person}{Cheng Guo}, \bibinfo{person}{Fei Sun}, \bibinfo{person}{Chao Li}, \bibinfo{person}{Andreas Pfadler}, \bibinfo{person}{Huan Zhao}, {and} \bibinfo{person}{Binqiang Zhao}.} \bibinfo{year}{2019}\natexlab{}.
\newblock \showarticletitle{POG: personalized outfit generation for fashion recommendation at Alibaba iFashion}. In \bibinfo{booktitle}{\emph{Proceedings of the 25th ACM SIGKDD international conference on knowledge discovery \& data mining}}. \bibinfo{pages}{2662--2670}.
\newblock


\bibitem[Ferraro et~al\mbox{.}(2018)]%
        {ferraro2018automatic}
\bibfield{author}{\bibinfo{person}{Andres Ferraro}, \bibinfo{person}{Dmitry Bogdanov}, \bibinfo{person}{Jisang Yoon}, \bibinfo{person}{KwangSeob Kim}, {and} \bibinfo{person}{Xavier Serra}.} \bibinfo{year}{2018}\natexlab{}.
\newblock \showarticletitle{Automatic playlist continuation using a hybrid recommender system combining features from text and audio}.
\newblock In \bibinfo{booktitle}{\emph{Proceedings of the ACM Recommender Systems Challenge 2018}}. \bibinfo{pages}{1--5}.
\newblock


\bibitem[He et~al\mbox{.}(2017)]%
        {he2017neural}
\bibfield{author}{\bibinfo{person}{Xiangnan He}, \bibinfo{person}{Lizi Liao}, \bibinfo{person}{Hanwang Zhang}, \bibinfo{person}{Liqiang Nie}, \bibinfo{person}{Xia Hu}, {and} \bibinfo{person}{Tat-Seng Chua}.} \bibinfo{year}{2017}\natexlab{}.
\newblock \showarticletitle{Neural Collaborative Filtering}. In \bibinfo{booktitle}{\emph{Proceedings of the 26th International Conference on World Wide Web}} (Perth, Australia) \emph{(\bibinfo{series}{WWW '17})}. \bibinfo{publisher}{International World Wide Web Conferences Steering Committee}, \bibinfo{address}{Republic and Canton of Geneva, CHE}, \bibinfo{pages}{173–182}.
\newblock
\showISBNx{9781450349130}
\href{https://doi.org/10.1145/3038912.3052569}{doi:\nolinkurl{10.1145/3038912.3052569}}


\bibitem[Hidasi et~al\mbox{.}(2015)]%
        {hidasi2015session}
\bibfield{author}{\bibinfo{person}{Bal{\'a}zs Hidasi}, \bibinfo{person}{Alexandros Karatzoglou}, \bibinfo{person}{Linas Baltrunas}, {and} \bibinfo{person}{Domonkos Tikk}.} \bibinfo{year}{2015}\natexlab{}.
\newblock \showarticletitle{Session-based recommendations with recurrent neural networks}.
\newblock \bibinfo{journal}{\emph{arXiv preprint arXiv:1511.06939}} (\bibinfo{year}{2015}).
\newblock


\bibitem[Hu et~al\mbox{.}(2025)]%
        {hu2025flashdlm}
\bibfield{author}{\bibinfo{person}{Zhanqiu Hu}, \bibinfo{person}{Jian Meng}, \bibinfo{person}{Yash Akhauri}, \bibinfo{person}{Mohamed~S Abdelfattah}, \bibinfo{person}{Jae-sun Seo}, \bibinfo{person}{Zhiru Zhang}, {and} \bibinfo{person}{Udit Gupta}.} \bibinfo{year}{2025}\natexlab{}.
\newblock \showarticletitle{FlashDLM: Accelerating Diffusion Language Model Inference via Efficient KV Caching and Guided Diffusion}.
\newblock \bibinfo{journal}{\emph{arXiv preprint arXiv:2505.21467}} (\bibinfo{year}{2025}).
\newblock


\bibitem[Jeon et~al\mbox{.}(2023)]%
        {jeon2023accurate}
\bibfield{author}{\bibinfo{person}{Hyunsik Jeon}, \bibinfo{person}{Jun-Gi Jang}, \bibinfo{person}{Taehun Kim}, {and} \bibinfo{person}{U Kang}.} \bibinfo{year}{2023}\natexlab{}.
\newblock \showarticletitle{Accurate bundle matching and generation via multitask learning with partially shared parameters}.
\newblock \bibinfo{journal}{\emph{Plos one}} \bibinfo{volume}{18}, \bibinfo{number}{3} (\bibinfo{year}{2023}), \bibinfo{pages}{e0280630}.
\newblock


\bibitem[Jin et~al\mbox{.}(2023)]%
        {jin2023language}
\bibfield{author}{\bibinfo{person}{Bowen Jin}, \bibinfo{person}{Hansi Zeng}, \bibinfo{person}{Guoyin Wang}, \bibinfo{person}{Xiusi Chen}, \bibinfo{person}{Tianxin Wei}, \bibinfo{person}{Ruirui Li}, \bibinfo{person}{Zhengyang Wang}, \bibinfo{person}{Zheng Li}, \bibinfo{person}{Yang Li}, \bibinfo{person}{Hanqing Lu}, {et~al\mbox{.}}} \bibinfo{year}{2023}\natexlab{}.
\newblock \showarticletitle{Language models as semantic indexers}.
\newblock \bibinfo{journal}{\emph{arXiv preprint arXiv:2310.07815}} (\bibinfo{year}{2023}).
\newblock


\bibitem[Kang and McAuley(2018)]%
        {kang2018self}
\bibfield{author}{\bibinfo{person}{Wang-Cheng Kang} {and} \bibinfo{person}{Julian McAuley}.} \bibinfo{year}{2018}\natexlab{}.
\newblock \showarticletitle{Self-attentive sequential recommendation}. In \bibinfo{booktitle}{\emph{2018 IEEE international conference on data mining (ICDM)}}. IEEE, \bibinfo{pages}{197--206}.
\newblock


\bibitem[Li et~al\mbox{.}(2020)]%
        {li2020time}
\bibfield{author}{\bibinfo{person}{Jiacheng Li}, \bibinfo{person}{Yujie Wang}, {and} \bibinfo{person}{Julian McAuley}.} \bibinfo{year}{2020}\natexlab{}.
\newblock \showarticletitle{Time interval aware self-attention for sequential recommendation}. In \bibinfo{booktitle}{\emph{Proceedings of the 13th international conference on web search and data mining}}. \bibinfo{pages}{322--330}.
\newblock


\bibitem[Li et~al\mbox{.}(2026)]%
        {li2026unigrec}
\bibfield{author}{\bibinfo{person}{Jialei Li}, \bibinfo{person}{Yang Zhang}, \bibinfo{person}{Yimeng Bai}, \bibinfo{person}{Shuai Zhu}, \bibinfo{person}{Ziqi Xue}, \bibinfo{person}{Xiaoyan Zhao}, \bibinfo{person}{Dingxian Wang}, \bibinfo{person}{Frank Yang}, \bibinfo{person}{Andrew Rabinovich}, {and} \bibinfo{person}{Xiangnan He}.} \bibinfo{year}{2026}\natexlab{}.
\newblock \showarticletitle{UniGRec: Unified Generative Recommendation with Soft Identifiers for End-to-End Optimization}.
\newblock \bibinfo{journal}{\emph{arXiv preprint arXiv:2601.17438}} (\bibinfo{year}{2026}).
\newblock


\bibitem[Liang et~al\mbox{.}(2026)]%
        {liang2026cd4lm}
\bibfield{author}{\bibinfo{person}{Yihao Liang}, \bibinfo{person}{Ze Wang}, \bibinfo{person}{Hao Chen}, \bibinfo{person}{Ximeng Sun}, \bibinfo{person}{Jialian Wu}, \bibinfo{person}{Xiaodong Yu}, \bibinfo{person}{Jiang Liu}, \bibinfo{person}{Emad Barsoum}, \bibinfo{person}{Zicheng Liu}, {and} \bibinfo{person}{Niraj~K Jha}.} \bibinfo{year}{2026}\natexlab{}.
\newblock \showarticletitle{CD4LM: Consistency Distillation and aDaptive Decoding for Diffusion Language Models}.
\newblock \bibinfo{journal}{\emph{arXiv preprint arXiv:2601.02236}} (\bibinfo{year}{2026}).
\newblock


\bibitem[Liu et~al\mbox{.}(2025)]%
        {liu2025fine}
\bibfield{author}{\bibinfo{person}{Xiaohao Liu}, \bibinfo{person}{Jie Wu}, \bibinfo{person}{Zhulin Tao}, \bibinfo{person}{Yunshan Ma}, \bibinfo{person}{Yinwei Wei}, {and} \bibinfo{person}{Tat-seng Chua}.} \bibinfo{year}{2025}\natexlab{}.
\newblock \showarticletitle{Fine-tuning multimodal large language models for product bundling}. In \bibinfo{booktitle}{\emph{Proceedings of the 31st ACM SIGKDD Conference on Knowledge Discovery and Data Mining V. 1}}. \bibinfo{pages}{848--858}.
\newblock


\bibitem[Liu et~al\mbox{.}(2026)]%
        {liu2026diffgrm}
\bibfield{author}{\bibinfo{person}{Zhao Liu}, \bibinfo{person}{Yichen Zhu}, \bibinfo{person}{Yiqing Yang}, \bibinfo{person}{Xiao Lv}, \bibinfo{person}{Guoping Tang}, \bibinfo{person}{Rui Huang}, \bibinfo{person}{Qiang Luo}, \bibinfo{person}{Ruiming Tang}, {and} \bibinfo{person}{Guorui Zhou}.} \bibinfo{year}{2026}\natexlab{}.
\newblock \showarticletitle{Diffgrm: Diffusion-based generative recommendation model}. In \bibinfo{booktitle}{\emph{Proceedings of the ACM Web Conference 2026}}. \bibinfo{pages}{5853--5864}.
\newblock


\bibitem[Ma et~al\mbox{.}(2024a)]%
        {ma2024multicbr}
\bibfield{author}{\bibinfo{person}{Yunshan Ma}, \bibinfo{person}{Yingzhi He}, \bibinfo{person}{Xiang Wang}, \bibinfo{person}{Yinwei Wei}, \bibinfo{person}{Xiaoyu Du}, \bibinfo{person}{Yuyangzi Fu}, {and} \bibinfo{person}{Tat-Seng Chua}.} \bibinfo{year}{2024}\natexlab{a}.
\newblock \showarticletitle{Multicbr: Multi-view contrastive learning for bundle recommendation}.
\newblock \bibinfo{journal}{\emph{ACM Transactions on Information Systems}} \bibinfo{volume}{42}, \bibinfo{number}{4} (\bibinfo{year}{2024}), \bibinfo{pages}{1--23}.
\newblock


\bibitem[Ma et~al\mbox{.}(2022)]%
        {ma2022crosscbr}
\bibfield{author}{\bibinfo{person}{Yunshan Ma}, \bibinfo{person}{Yingzhi He}, \bibinfo{person}{An Zhang}, \bibinfo{person}{Xiang Wang}, {and} \bibinfo{person}{Tat-Seng Chua}.} \bibinfo{year}{2022}\natexlab{}.
\newblock \showarticletitle{CrossCBR: cross-view contrastive learning for bundle recommendation}. In \bibinfo{booktitle}{\emph{Proceedings of the 28th ACM SIGKDD conference on knowledge discovery and data mining}}. \bibinfo{pages}{1233--1241}.
\newblock


\bibitem[Ma et~al\mbox{.}(2024b)]%
        {ma2024leveraging}
\bibfield{author}{\bibinfo{person}{Yunshan Ma}, \bibinfo{person}{Xiaohao Liu}, \bibinfo{person}{Yinwei Wei}, \bibinfo{person}{Zhulin Tao}, \bibinfo{person}{Xiang Wang}, {and} \bibinfo{person}{Tat-Seng Chua}.} \bibinfo{year}{2024}\natexlab{b}.
\newblock \showarticletitle{Leveraging multimodal features and item-level user feedback for bundle construction}. In \bibinfo{booktitle}{\emph{Proceedings of the 17th ACM International Conference on Web Search and Data Mining}}. \bibinfo{pages}{510--519}.
\newblock


\bibitem[Ni et~al\mbox{.}(2026)]%
        {ni2026flexibility}
\bibfield{author}{\bibinfo{person}{Zanlin Ni}, \bibinfo{person}{Shenzhi Wang}, \bibinfo{person}{Yang Yue}, \bibinfo{person}{Tianyu Yu}, \bibinfo{person}{Weilin Zhao}, \bibinfo{person}{Yeguo Hua}, \bibinfo{person}{Tianyi Chen}, \bibinfo{person}{Jun Song}, \bibinfo{person}{Cheng Yu}, \bibinfo{person}{Bo Zheng}, {et~al\mbox{.}}} \bibinfo{year}{2026}\natexlab{}.
\newblock \showarticletitle{The Flexibility Trap: Rethinking the Value of Arbitrary Order in Diffusion Language Models}.
\newblock  (\bibinfo{year}{2026}).
\newblock


\bibitem[Nie et~al\mbox{.}(2026)]%
        {nie2026large}
\bibfield{author}{\bibinfo{person}{Shen Nie}, \bibinfo{person}{Fengqi Zhu}, \bibinfo{person}{Zebin You}, \bibinfo{person}{Xiaolu Zhang}, \bibinfo{person}{Jingyang Ou}, \bibinfo{person}{Jun Hu}, \bibinfo{person}{Jun Zhou}, \bibinfo{person}{Yankai Lin}, \bibinfo{person}{Ji-Rong Wen}, {and} \bibinfo{person}{Chongxuan Li}.} \bibinfo{year}{2026}\natexlab{}.
\newblock \showarticletitle{Large language diffusion models}.
\newblock \bibinfo{journal}{\emph{Advances in Neural Information Processing Systems}}  \bibinfo{volume}{38} (\bibinfo{year}{2026}), \bibinfo{pages}{50608--50646}.
\newblock


\bibitem[Petrov and Macdonald(2024)]%
        {petrov2024recjpq}
\bibfield{author}{\bibinfo{person}{Aleksandr~V Petrov} {and} \bibinfo{person}{Craig Macdonald}.} \bibinfo{year}{2024}\natexlab{}.
\newblock \showarticletitle{RecJPQ: training large-catalogue sequential recommenders}. In \bibinfo{booktitle}{\emph{Proceedings of the 17th ACM International Conference on Web Search and Data Mining}}. \bibinfo{pages}{538--547}.
\newblock


\bibitem[Qian et~al\mbox{.}(2026)]%
        {qian2026d3llm}
\bibfield{author}{\bibinfo{person}{Yu-Yang Qian}, \bibinfo{person}{Junda Su}, \bibinfo{person}{Lanxiang Hu}, \bibinfo{person}{Peiyuan Zhang}, \bibinfo{person}{Zhijie Deng}, \bibinfo{person}{Peng Zhao}, {and} \bibinfo{person}{Hao Zhang}.} \bibinfo{year}{2026}\natexlab{}.
\newblock \showarticletitle{d3LLM: Ultra-Fast Diffusion LLM using Pseudo-Trajectory Distillation}.
\newblock \bibinfo{journal}{\emph{arXiv preprint arXiv:2601.07568}} (\bibinfo{year}{2026}).
\newblock


\bibitem[Rajput et~al\mbox{.}({[n.\,d.]})]%
        {rajputrecommender}
\bibfield{author}{\bibinfo{person}{Shashank Rajput}, \bibinfo{person}{Nikhil Mehta}, \bibinfo{person}{Anima Singh}, \bibinfo{person}{Raghunandan Keshavan}, \bibinfo{person}{Trung Vu}, \bibinfo{person}{Lukasz Heldt}, \bibinfo{person}{Lichan Hong}, \bibinfo{person}{Yi Tay}, \bibinfo{person}{Vinh~Q Tran}, \bibinfo{person}{Jonah Samost}, {et~al\mbox{.}}} \bibinfo{year}{[n.\,d.]}\natexlab{}.
\newblock \showarticletitle{Recommender Systems with Generative Retrieval}.
\newblock  (\bibinfo{year}{[n.\,d.]}).
\newblock


\bibitem[Sahoo et~al\mbox{.}(2024)]%
        {sahoo2024simple}
\bibfield{author}{\bibinfo{person}{Subham~Sekhar Sahoo}, \bibinfo{person}{Marianne Arriola}, \bibinfo{person}{Yair Schiff}, \bibinfo{person}{Aaron Gokaslan}, \bibinfo{person}{Edgar Marroquin}, \bibinfo{person}{Justin~T Chiu}, \bibinfo{person}{Alexander Rush}, {and} \bibinfo{person}{Volodymyr Kuleshov}.} \bibinfo{year}{2024}\natexlab{}.
\newblock \showarticletitle{Simple and effective masked diffusion language models}. In \bibinfo{booktitle}{\emph{Proceedings of the 38th International Conference on Neural Information Processing Systems}}. \bibinfo{pages}{130136--130184}.
\newblock


\bibitem[Shao et~al\mbox{.}(2024)]%
        {shao2024deepseekmath}
\bibfield{author}{\bibinfo{person}{Zhihong Shao}, \bibinfo{person}{Peiyi Wang}, \bibinfo{person}{Qihao Zhu}, \bibinfo{person}{Runxin Xu}, \bibinfo{person}{Junxiao Song}, \bibinfo{person}{Xiao Bi}, \bibinfo{person}{Haowei Zhang}, \bibinfo{person}{Mingchuan Zhang}, \bibinfo{person}{YK Li}, \bibinfo{person}{Yang Wu}, {et~al\mbox{.}}} \bibinfo{year}{2024}\natexlab{}.
\newblock \showarticletitle{Deepseekmath: Pushing the limits of mathematical reasoning in open language models}.
\newblock \bibinfo{journal}{\emph{arXiv preprint arXiv:2402.03300}} (\bibinfo{year}{2024}).
\newblock


\bibitem[Si et~al\mbox{.}(2024)]%
        {si2024generative}
\bibfield{author}{\bibinfo{person}{Zihua Si}, \bibinfo{person}{Zhongxiang Sun}, \bibinfo{person}{Jiale Chen}, \bibinfo{person}{Guozhang Chen}, \bibinfo{person}{Xiaoxue Zang}, \bibinfo{person}{Kai Zheng}, \bibinfo{person}{Yang Song}, \bibinfo{person}{Xiao Zhang}, \bibinfo{person}{Jun Xu}, {and} \bibinfo{person}{Kun Gai}.} \bibinfo{year}{2024}\natexlab{}.
\newblock \showarticletitle{Generative retrieval with semantic tree-structured identifiers and contrastive learning}. In \bibinfo{booktitle}{\emph{Proceedings of the 2024 Annual International ACM SIGIR Conference on Research and Development in Information Retrieval in the Asia Pacific Region}}. \bibinfo{pages}{154--163}.
\newblock


\bibitem[Song et~al\mbox{.}(2019)]%
        {song2019gpbpr}
\bibfield{author}{\bibinfo{person}{Xuemeng Song}, \bibinfo{person}{Xianjing Han}, \bibinfo{person}{Yunkai Li}, \bibinfo{person}{Jingyuan Chen}, \bibinfo{person}{Xin-Shun Xu}, {and} \bibinfo{person}{Liqiang Nie}.} \bibinfo{year}{2019}\natexlab{}.
\newblock \showarticletitle{GP-BPR: Personalized Compatibility Modeling for Clothing Matching}. In \bibinfo{booktitle}{\emph{Proceedings of the 27th ACM International Conference on Multimedia}} (Nice, France) \emph{(\bibinfo{series}{MM '19})}. \bibinfo{publisher}{Association for Computing Machinery}, \bibinfo{address}{New York, NY, USA}, \bibinfo{pages}{320–328}.
\newblock
\showISBNx{9781450368896}
\href{https://doi.org/10.1145/3343031.3350956}{doi:\nolinkurl{10.1145/3343031.3350956}}


\bibitem[Sun et~al\mbox{.}(2019)]%
        {sun2019bert4rec}
\bibfield{author}{\bibinfo{person}{Fei Sun}, \bibinfo{person}{Jun Liu}, \bibinfo{person}{Jian Wu}, \bibinfo{person}{Changhua Pei}, \bibinfo{person}{Xiao Lin}, \bibinfo{person}{Wenwu Ou}, {and} \bibinfo{person}{Peng Jiang}.} \bibinfo{year}{2019}\natexlab{}.
\newblock \showarticletitle{BERT4Rec: Sequential recommendation with bidirectional encoder representations from transformer}. In \bibinfo{booktitle}{\emph{Proceedings of the 28th ACM international conference on information and knowledge management}}. \bibinfo{pages}{1441--1450}.
\newblock


\bibitem[Tang and Wang(2018)]%
        {tang2018personalized}
\bibfield{author}{\bibinfo{person}{Jiaxi Tang} {and} \bibinfo{person}{Ke Wang}.} \bibinfo{year}{2018}\natexlab{}.
\newblock \showarticletitle{Personalized top-n sequential recommendation via convolutional sequence embedding}. In \bibinfo{booktitle}{\emph{Proceedings of the eleventh ACM international conference on web search and data mining}}. \bibinfo{pages}{565--573}.
\newblock


\bibitem[Tu et~al\mbox{.}(2026)]%
        {tu2026discrete}
\bibfield{author}{\bibinfo{person}{Teng Tu}, \bibinfo{person}{Ai Li}, \bibinfo{person}{Yunshan Ma}, \bibinfo{person}{Shuo Xu}, \bibinfo{person}{Xiaohao Liu}, \bibinfo{person}{Haokai Ma}, \bibinfo{person}{Liang Pang}, {and} \bibinfo{person}{Tat-Seng Chua}.} \bibinfo{year}{2026}\natexlab{}.
\newblock \showarticletitle{Discrete diffusion for bundle construction}. In \bibinfo{booktitle}{\emph{The Fourteenth International Conference on Learning Representations}}.
\newblock


\bibitem[Vall et~al\mbox{.}(2017)]%
        {vall2017music}
\bibfield{author}{\bibinfo{person}{Andreu Vall}, \bibinfo{person}{Hamid Eghbal-Zadeh}, \bibinfo{person}{Matthias Dorfer}, \bibinfo{person}{Markus Schedl}, {and} \bibinfo{person}{Gerhard Widmer}.} \bibinfo{year}{2017}\natexlab{}.
\newblock \showarticletitle{Music playlist continuation by learning from hand-curated examples and song features: Alleviating the cold-start problem for rare and out-of-set songs}. In \bibinfo{booktitle}{\emph{Proceedings of the 2nd workshop on deep learning for recommender systems}}. \bibinfo{pages}{46--54}.
\newblock


\bibitem[Wang et~al\mbox{.}(2024)]%
        {wang2024eager}
\bibfield{author}{\bibinfo{person}{Ye Wang}, \bibinfo{person}{Jiahao Xun}, \bibinfo{person}{Minjie Hong}, \bibinfo{person}{Jieming Zhu}, \bibinfo{person}{Tao Jin}, \bibinfo{person}{Wang Lin}, \bibinfo{person}{Haoyuan Li}, \bibinfo{person}{Linjun Li}, \bibinfo{person}{Yan Xia}, \bibinfo{person}{Zhou Zhao}, {et~al\mbox{.}}} \bibinfo{year}{2024}\natexlab{}.
\newblock \showarticletitle{Eager: Two-stream generative recommender with behavior-semantic collaboration}. In \bibinfo{booktitle}{\emph{Proceedings of the 30th ACM SIGKDD Conference on Knowledge Discovery and Data Mining}}. \bibinfo{pages}{3245--3254}.
\newblock


\bibitem[Wu et~al\mbox{.}(2025)]%
        {wu2025fast}
\bibfield{author}{\bibinfo{person}{Chengyue Wu}, \bibinfo{person}{Hao Zhang}, \bibinfo{person}{Shuchen Xue}, \bibinfo{person}{Zhijian Liu}, \bibinfo{person}{Shizhe Diao}, \bibinfo{person}{Ligeng Zhu}, \bibinfo{person}{Ping Luo}, \bibinfo{person}{Song Han}, {and} \bibinfo{person}{Enze Xie}.} \bibinfo{year}{2025}\natexlab{}.
\newblock \showarticletitle{Fast-dllm: Training-free acceleration of diffusion llm by enabling kv cache and parallel decoding}.
\newblock \bibinfo{journal}{\emph{arXiv preprint arXiv:2505.22618}} (\bibinfo{year}{2025}).
\newblock


\bibitem[Yuan et~al\mbox{.}(2019)]%
        {yuan2019simple}
\bibfield{author}{\bibinfo{person}{Fajie Yuan}, \bibinfo{person}{Alexandros Karatzoglou}, \bibinfo{person}{Ioannis Arapakis}, \bibinfo{person}{Joemon~M Jose}, {and} \bibinfo{person}{Xiangnan He}.} \bibinfo{year}{2019}\natexlab{}.
\newblock \showarticletitle{A simple convolutional generative network for next item recommendation}. In \bibinfo{booktitle}{\emph{Proceedings of the twelfth ACM international conference on web search and data mining}}. \bibinfo{pages}{582--590}.
\newblock


\bibitem[Zhai et~al\mbox{.}(2024)]%
        {zhai2024actions}
\bibfield{author}{\bibinfo{person}{Jiaqi Zhai}, \bibinfo{person}{Lucy Liao}, \bibinfo{person}{Xing Liu}, \bibinfo{person}{Yueming Wang}, \bibinfo{person}{Rui Li}, \bibinfo{person}{Xuan Cao}, \bibinfo{person}{Leon Gao}, \bibinfo{person}{Zhaojie Gong}, \bibinfo{person}{Fangda Gu}, \bibinfo{person}{Michael He}, {et~al\mbox{.}}} \bibinfo{year}{2024}\natexlab{}.
\newblock \showarticletitle{Actions speak louder than words: Trillion-parameter sequential transducers for generative recommendations}.
\newblock \bibinfo{journal}{\emph{arXiv preprint arXiv:2402.17152}} (\bibinfo{year}{2024}).
\newblock


\bibitem[Zhang et~al\mbox{.}(2026)]%
        {zhang2026few}
\bibfield{author}{\bibinfo{person}{Tunyu Zhang}, \bibinfo{person}{Xinxi Zhang}, \bibinfo{person}{Ligong Han}, \bibinfo{person}{Haizhou Shi}, \bibinfo{person}{Xiaoxiao He}, \bibinfo{person}{Zhuowei Li}, \bibinfo{person}{Hao Wang}, \bibinfo{person}{Kai Xu}, \bibinfo{person}{Akash Srivastava}, \bibinfo{person}{Chengzhi Mao}, {et~al\mbox{.}}} \bibinfo{year}{2026}\natexlab{}.
\newblock \showarticletitle{Few-Step Diffusion Language Models via Trajectory Self-Distillation}.
\newblock \bibinfo{journal}{\emph{arXiv preprint arXiv:2602.12262}} (\bibinfo{year}{2026}).
\newblock


\bibitem[Zhang et~al\mbox{.}(2019)]%
        {zhang2019feature}
\bibfield{author}{\bibinfo{person}{Tingting Zhang}, \bibinfo{person}{Pengpeng Zhao}, \bibinfo{person}{Yanchi Liu}, \bibinfo{person}{Victor~S Sheng}, \bibinfo{person}{Jiajie Xu}, \bibinfo{person}{Deqing Wang}, \bibinfo{person}{Guanfeng Liu}, \bibinfo{person}{Xiaofang Zhou}, {et~al\mbox{.}}} \bibinfo{year}{2019}\natexlab{}.
\newblock \showarticletitle{Feature-level deeper self-attention network for sequential recommendation.}. In \bibinfo{booktitle}{\emph{IJCAI}}. \bibinfo{pages}{4320--4326}.
\newblock


\bibitem[Zheng et~al\mbox{.}(2024)]%
        {zheng2024adapting}
\bibfield{author}{\bibinfo{person}{Bowen Zheng}, \bibinfo{person}{Yupeng Hou}, \bibinfo{person}{Hongyu Lu}, \bibinfo{person}{Yu Chen}, \bibinfo{person}{Wayne~Xin Zhao}, \bibinfo{person}{Ming Chen}, {and} \bibinfo{person}{Ji-Rong Wen}.} \bibinfo{year}{2024}\natexlab{}.
\newblock \showarticletitle{Adapting large language models by integrating collaborative semantics for recommendation}. In \bibinfo{booktitle}{\emph{2024 IEEE 40th International Conference on Data Engineering (ICDE)}}. IEEE, \bibinfo{pages}{1435--1448}.
\newblock


\bibitem[Zhou et~al\mbox{.}(2020)]%
        {zhou2020s3}
\bibfield{author}{\bibinfo{person}{Kun Zhou}, \bibinfo{person}{Hui Wang}, \bibinfo{person}{Wayne~Xin Zhao}, \bibinfo{person}{Yutao Zhu}, \bibinfo{person}{Sirui Wang}, \bibinfo{person}{Fuzheng Zhang}, \bibinfo{person}{Zhongyuan Wang}, {and} \bibinfo{person}{Ji-Rong Wen}.} \bibinfo{year}{2020}\natexlab{}.
\newblock \showarticletitle{S3-rec: Self-supervised learning for sequential recommendation with mutual information maximization}. In \bibinfo{booktitle}{\emph{Proceedings of the 29th ACM international conference on information \& knowledge management}}. \bibinfo{pages}{1893--1902}.
\newblock


\bibitem[Zhu et~al\mbox{.}(2014)]%
        {zhu2014bundle}
\bibfield{author}{\bibinfo{person}{Tao Zhu}, \bibinfo{person}{Patrick Harrington}, \bibinfo{person}{Junjun Li}, {and} \bibinfo{person}{Lei Tang}.} \bibinfo{year}{2014}\natexlab{}.
\newblock \showarticletitle{Bundle recommendation in ecommerce}. In \bibinfo{booktitle}{\emph{Proceedings of the 37th international ACM SIGIR conference on Research \& development in information retrieval}}. \bibinfo{pages}{657--666}.
\newblock


\end{thebibliography}

\section{Appendix}
\subsection{Dataset and Tokenization Details}
\label{sec:dataset_detail}
\paragraph{Dataset Statistics.}
Spotify contains 254,155 tracks, and POG contains 31,217 fashion products. IQON3000 contains 715,883 fashion items after preprocessing.
\paragraph{Preprocessing.}
Let $K$ denote the fixed bundle length for each benchmark. Bundles containing fewer than $K$ items are removed, while longer bundles are truncated to $K$ items. Spotify and POG follow the released partitions. For IQON3000, we drop outfits with missing item images, randomly shuffle the items within each outfit, and randomly split outfits into training, validation, and test sets at a 7:1:2 ratio. All partitions contain disjoint bundles. For evaluation, the first $\lceil K/2\rceil$ items form the observed bundle $\mathbf{c}$, while the remaining $N=\lfloor K/2\rfloor$ items form the ground truth target bundle $\mathcal{B}_{\star}$.

\paragraph{Training Augmentation.}
For each training bundle
$$\mathcal{B}=[b^{(1)},\ldots,b^{(|\mathcal{B}|)}]$$
before truncation, we generate
$\lfloor r|\mathcal{B}|\rfloor$
augmented samples. Each sample first swaps one pair of adjacent items,
$b^{(j)}$ and $b^{(j+1)}$, where
$j\sim\mathcal{U}\{1,\ldots,|\mathcal{B}|-1\}$,
and then selects a random contiguous subsequence of $K$ items. We set $r=0.4$ for Spotify and $r=0.8$ for POG and IQON3000. Validation and test bundles are not augmented.

\paragraph{Semantic Identifier Tokenization.}
Each item $i$ has a fixed CLHE representation
$\mathbf{e}_i\in\mathbb{R}^{64}$,
taken from the CLHE release for Spotify and POG and trained with the official CLHE code on image and text features for IQON3000,
and is assigned a four token SID
$\mathbf{z}_i=[z_{i,1},\ldots,z_{i,4}]$.
The first three tokens are obtained using residual vector quantization. Let
$\{\mathbf{q}_{m,1},\ldots,\mathbf{q}_{m,C}\}$
denote the codebook at semantic level $m$, where $C=256$ for Spotify and IQON3000 and $C=128$ for POG. The quantization process is defined as

\begin{equation}
\begin{aligned}
\mathbf{r}_{i,0}
&=
\mathbf{e}_i,
\qquad
z_{i,m}
=
\operatorname*{arg\,min}_{z\in\{1,\ldots,C\}}
\left\|
\mathbf{r}_{i,m-1}
-
\mathbf{q}_{m,z}
\right\|^{2},
\\
\mathbf{r}_{i,m}
&=
\mathbf{r}_{i,m-1}
-
\mathbf{q}_{m,z_{i,m}},
\qquad
m\in\{1,2,3\}.
\end{aligned}
\end{equation}

The fourth token distinguishes items that share the same first three tokens. Items with an identical prefix
$[z_{i,1},z_{i,2},z_{i,3}]$
are assigned distinct values of $z_{i,4}$ according to their catalog order, ensuring that each complete SID uniquely identifies one item. The maximum number of collisions is 74 for Spotify, 53 for POG, and 171 for IQON3000, all below $C$.

\paragraph{Sequence Layout.}
A bundle of $K$ items is serialized as

\begin{equation}
[\mathrm{BOS}],
[\mathrm{BOI},z_{1,1},\ldots,z_{1,4}],
\ldots,
[\mathrm{BOI},z_{K,1},\ldots,z_{K,4}],
[\mathrm{EOS}],
\end{equation}

resulting in a sequence of length $5K+2$. Tokens from semantic level $m$ are mapped to disjoint vocabulary ranges using

\begin{equation}
\operatorname{tok}(z_{i,m},m)
=
z_{i,m}
+
(m-1)C.
\end{equation}

The resulting vocabulary contains $4C$ SID tokens and three special tokens, with an additional $[\mathrm{MASK}]$ token used by the diffusion model. During evaluation, the observed bundle $\mathbf{c}$ and the target bundle $\mathcal{B}_{\star}$ are serialized as two consecutive segments, each with its own $[\mathrm{BOS}]$ and $[\mathrm{EOS}]$; the SIDs of $\mathbf{c}$ remain visible, while all SID positions of the target segment are replaced with $[\mathrm{MASK}]$.

\subsection{Training Details of \name}
\label{sec:training_detail}

All experiments run on a single NVIDIA A100-80GB GPU. For Spotify and POG we pretrain the backbone
from scratch with the identical DDBC recipe. The GRPO stage initializes from the
pretrained backbone and selects its checkpoint by mean validation rollout reward;
the distillation stage initializes from the teacher model trained from GRPO policy. Table~\ref{tab:hparams} lists all hyperparameters.

\begin{table}[h]
\centering
\caption{Training hyperparameters of \name.}
\label{tab:hparams}
\small
\resizebox{\columnwidth}{!}{%
\begin{tabular}{lccc}
\toprule
 & Pretraining & GRPO & Distillation \\
\midrule
Optimizer & AdamW & AdamW & AdamW \\
Learning rate & $3\times10^{-4}$ & $1\times10^{-5}$ & $2\times10^{-5}$ \\
Weight decay & 0 & 0 & 0.01 \\
LR schedule & warmup 500, constant & constant & constant \\
Training steps & 20{,}000 & 1{,}000 & 5{,}000 \\
Batch size & 512 & 512 rollouts (32 bundles $\times$ $G{=}16$) & 512 \\
Gradient clipping & 1.0 & 1.0 & 1.0 \\
Precision & bf16 & bf16 & bf16 \\
EMA & 0.9999 & -- & -- \\
Validation interval & 32 & 50 & 500 \\
Selection metric & val.\ NLL & val.\ rollout reward & fixed step \\
\midrule
Temperatures $\boldsymbol{\tau}$ & -- & $(0.6, 0.2, 0.2, 0.2)$ & inherited \\
Reward weights $(w_1, w_2)$ & -- & $(0.2, 0.5)$ & -- \\
Group size $G$ & -- & 16 & -- \\
Mask-ratio cap $t_{\max}$ & -- & -- & 0.8 \\
Trajectories per bundle & -- & -- & 1 \\
\bottomrule
\end{tabular}%
}
\end{table}

\subsection{Evaluation Metric Details}
\label{sec:metric_detail}

Let
\begin{equation}
\mathcal{B}_{\star}
=
\left\{
b_{\star}^{(1)},\ldots,b_{\star}^{(N)}
\right\},
\qquad
\widehat{\mathcal{B}}
=
\left\{
\hat{b}^{(1)},\ldots,\hat{b}^{(N)}
\right\}
\end{equation}
denote the ground truth target bundle and the predicted bundle, respectively. Under the equal input and target partition, every method generates exactly $N$ items. Since bundle construction produces an unordered set, all metrics are independent of item order and are computed for each bundle before being averaged over the test set.

\paragraph{F1.}
Precision and recall are defined as

\begin{equation}
P
=
\frac{
\left|
\widehat{\mathcal{B}}
\cap
\mathcal{B}_{\star}
\right|
}{
\left|
\widehat{\mathcal{B}}
\right|
},
\qquad
R
=
\frac{
\left|
\widehat{\mathcal{B}}
\cap
\mathcal{B}_{\star}
\right|
}{
\left|
\mathcal{B}_{\star}
\right|
},
\end{equation}

and the F1 score is

\begin{equation}
\mathrm{F1}
=
\frac{2PR}{P+R}.
\end{equation}

Since
$\left|\widehat{\mathcal{B}}\right|
=
\left|\mathcal{B}_{\star}\right|
=
N$
and the predicted items are distinct, precision, recall, and F1 are equivalent.

\paragraph{Jaccard Similarity.}

\begin{equation}
\mathrm{Jaccard}
=
\frac{
\left|
\widehat{\mathcal{B}}
\cap
\mathcal{B}_{\star}
\right|
}{
\left|
\widehat{\mathcal{B}}
\cup
\mathcal{B}_{\star}
\right|
}.
\end{equation}

\paragraph{Optimal Assignment Similarity.}

Let $\mathbf{e}_i\in\mathbb{R}^{d}$ denote the fixed pretrained representation of item $i$. The cosine similarity between items $i$ and $j$ is

\begin{equation}
s(i,j)
=
\frac{
\mathbf{e}_{i}^{\top}\mathbf{e}_{j}
}{
\left\|\mathbf{e}_{i}\right\|
\left\|\mathbf{e}_{j}\right\|
}.
\end{equation}

Optimal assignment similarity measures the mean similarity under the best one to one matching between the predicted and ground truth bundles:

\begin{equation}
\mathrm{OAS}
=
\max_{\pi\in\Pi_N}
\frac{1}{N}
\sum_{n=1}^{N}
s
\left(
\hat{b}^{(\pi(n))},
b_{\star}^{(n)}
\right),
\end{equation}

where $\Pi_N$ denotes the set of permutations over $\{1,\ldots,N\}$. The optimal assignment is obtained using the Hungarian algorithm. Unlike F1 and Jaccard similarity, which recognize only exact item matches, OAS also rewards predictions that are semantically close to the target items in the pretrained representation space.

\subsection{Diversity Analysis}
\label{sec:diversity_analysis}

To complement the pairwise cosine similarity analysis in Figure~\ref{fig:cross_item_similarity}, we evaluate catalog coverage and cluster entropy on the three Spotify settings. Catalog coverage measures the percentage of catalog items recommended at least once across the test set. Cluster entropy measures the dispersion of recommended items across semantic clusters, with higher values indicating a broader and more balanced distribution.

\begin{figure}
    \centering
    \subfloat[]{\includegraphics[width=1\columnwidth]{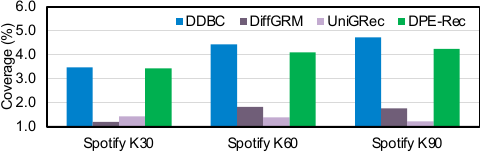}
    \label{fig:diversity_coverage}
    }
    \vspace{0pt}
    \subfloat[]{\includegraphics[width=1\columnwidth]{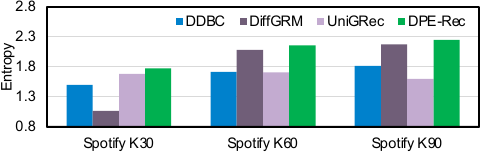}
    \label{fig:diversity_entropy}
    }
    \caption{
    (a) Catalog coverage on Spotify;
    (b) Cluster entropy on Spotify.
    }
    \label{fig:diversity}
\end{figure}

Figure~\ref{fig:diversity} shows that \name achieves the highest cluster entropy across all three settings and 2.2--3.4$\times$ the catalog coverage of DiffGRM and UniGRec. Compared with DDBC, \name obtains higher cluster entropy despite lower catalog coverage, highlighting that catalog exposure and semantic diversity capture complementary aspects of recommendation. Together with its lower pairwise cosine similarity, these results are consistent with fork-aware GRPO preserving exploration across alternative semantic branches.

\subsection{Fine-Grained Ablation Study}
\label{sec:detail_ablation}

We further examine the hierarchical reward design and GRPO group size. The reward weights $w_1$ and $w_2$ control agreement at the two-token SID-prefix level and the complete-SID level, respectively. We vary each parameter independently around the default configuration $(w_1,w_2,G)=(0.2,0.5,16)$. Setting a reward weight to zero removes the corresponding term. For the group-size sweep, we fix the total number of rollouts per update at 512. We report F1 scores separately on Spotify$_{30}$, POG, and IQON3000.

\begin{figure}
    \centering
    \includegraphics[width=1\columnwidth]{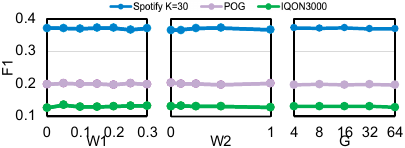}
    \vspace{-15pt}
    \caption{Sensitivity to reward weights and GRPO group size. Each parameter is varied independently while the remaining parameters retain their default values.
   }
    \label{fig:hyperPara}
    \vspace{-15pt}
\end{figure}

\paragraph{Reward design.}
Figure~\ref{fig:hyperPara} shows that recommendation quality is relatively insensitive to the reward weights within the tested ranges, reducing the need for careful reward tuning.

\paragraph{Group size.}
Figure~\ref{fig:hyperPara} shows that under a fixed rollout budget, larger groups provide no consistent quality benefit, suggesting that effective post-training does not require large rollout groups.

\subsection{Training Cost Analysis}
\label{sec:training_cost}

Table~\ref{tab:training_cost} reports the training cost of each stage on a single NVIDIA A100-80GB GPU. GRPO accounts for most of the additional training cost on Spotify, consistent with the overhead of autoregressive rollouts for longer bundles. These costs are incurred during post-training, while the resulting inference savings apply to every subsequently generated bundle.

\begin{table}[htbp]
\centering
\caption{Training cost in GPU-hours on a single NVIDIA A100-80GB GPU. Additional cost is the sum of GRPO and PTD training time.}
\label{tab:training_cost}
\setlength{\tabcolsep}{4pt}
\resizebox{\columnwidth}{!}{%
\begin{tabular}{lccccc}
\toprule
Stage & Spotify$_{30}$ & Spotify$_{60}$ & Spotify$_{90}$ & POG & IQON3000 \\
\midrule
Base DLM & 0.6 & 0.9 & 1.2 & 0.3 & 0.2 \\
GRPO     & 1.2 & 4.5 & 10.0 & 0.1 & 0.2 \\
PTD      & 0.4 & 0.9 & 1.3 & 0.1 & 0.1 \\
\midrule
Additional cost & 1.6 & 5.4 & 11.3 & 0.2 & 0.3 \\
\bottomrule
\end{tabular}%
}
\end{table}

To relate this overhead to inference benefits, we estimate the number of generated bundles required to offset the additional GRPO and PTD training time:
\[
N_{\mathrm{amort}}
=
\frac{T_{\mathrm{GRPO}}+T_{\mathrm{PTD}}}
{L_{\mathrm{DDBC}}-L_{\mathrm{Fast}}},
\]
where $T$ denotes the reported training time and $L$ denotes per-bundle inference latency, with all times converted to seconds. Using DDBC's original decoding configuration as the reference and the latencies in Table~\ref{tab:speed_up}, the additional training cost is amortized after approximately 1.3K--5.0K generated bundles across the five evaluation settings.

\subsection{Sensitivity to Baseline Selection}
\label{sec:baseline_selection}

Some baselines achieve relatively low recommendation quality, potentially inflating average relative improvements. To assess whether the reported gains depend heavily on these weaker baselines, we recompute the results in Tables~\ref{tab:main_results} and~\ref{tab:speed_up} over five representative baselines: FDSA, HSTU, UniGRec, DiffGRM, and DDBC. This subset retains competitive discriminative models and all three generative baselines, excluding GRU4Rec, SASRec, BERT4Rec, and RecJPQ.

Relative quality improvements are computed against each retained baseline and averaged equally across baselines, evaluation settings, and metrics. Both \name and \name-Fast achieve approximately 26.9\% average improvement under this comparison. \name-Fast also achieves a 15.6$\times$ average speedup, with latency ratios averaged over the same five baselines and five evaluation settings.

\end{document}